\documentclass[nofootinbib,twocolumn,showpacs,preprintnumbers,amsmath,amssymb,aps,prd]{revtex4}
\usepackage{graphicx}
\begin{document}
\title{Geometrical optics analysis of the short-time stability properties of the Einstein
  evolution equations}
\author{R. O'Shaughnessy}
\email{oshaughn@caltech.edu}
\affiliation{Theoretical Astrophysics, California Institute of Technology, Pasadena, CA 91125}
\date{Received ?? Month 2003}
\begin{abstract}
Many alternative formulations of Einstein's evolution have lately been   examined, 
in an effort to
discover one which yields slow growth of constraint-violating errors. 
In this paper, rather than directly  search for well-behaved formulations,
we instead develop analytic tools to discover which formulations are particularly ill-behaved.
Specifically, we examine the growth of  approximate (geometric-optics) solutions, studied only
in the future 
domain of dependence of the initial data slice (e.g. we study transients).
By evaluating the amplification of transients a given formulation will produce, we may
therefore eliminate from consideration the most pathological formulations (e.g. those with
numerically-unacceptable amplification).  This technique has 
the potential to provide surprisingly tight constraints on the set of formulations one can
safely apply.
 To illustrate the application of these techniques to practical examples, we apply our technique to
the 2-parameter family of evolution equations proposed by Kidder, Scheel, and Teukolsky,
focusing in particular on flat space (in Rindler coordinates) and Schwarzchild (in
Painleve-Gullstrand coordinates).
\end{abstract}
\maketitle

\section{Introduction}
Recently developed numerical codes offer the possibility of extremely accurate and
computationally efficient evolutions of  Einstein's evolution equations in vacuum \cite{KST}.   
To take full advantage of these new techniques to perform an unconstrained evolution of initial
data and boundary conditions, we must address an unpleasant fact: many choices for evolution
equations and and boundary conditions permit ill-behaved, unphysical solutions (e.g. growing,
constraint-violating solutions) near physical solutions.    

By way of example, when
Kidder, Scheel, and Teukolsky (KST)  evolved  a single static Schwarzchild hole as a test
case, they 
found evidence suggesting that their evolution equations and boundary conditions, when
linearized about a Schwarzchild 
background, admitted growing, constraint-violating \emph{eigenmodes} \cite{KST}
\cite{LSenNorms}.  These eigenmodes were 
excited by generic initial data (i.e. roundoff error); grew to significant magnitude; and were
directly correlated with the time their code crashed.
As this example demonstrates, the existence and growth of  ill-behaved solutions  limit the
length  
of time a given numerical simulation can be trusted -- or even run.

For this reason, some researchers have explored the analytic properties of various formulations
of Einstein's equations
\cite{LSenNorms}   \cite{paADM}  \cite{paShinkaiReview} 
\cite{paWeakHyperbolicBad1} \cite{paFR}
\cite{paBSSN} \cite{paEarlyWorkAddConstraints}  
and boundary conditions   
\cite{bcWellPosed1} \cite{bcWellPosed2}  \cite{bcWellPosed3} \cite{bcFrittelli}
used in numerical relativity, searching for ways to understand and control these
undesirable perturbations.

In this paper, we discuss one particular type of undesirable perturbation: short-wavelength,
transient wave packets.  
[For the purposes of this paper, a
transient will be any solution defined in the future domain of dependence of the initial data
slice.  Depending on the
boundary conditions, the solution may or may not extend farther in time, outside the future
domain of dependence.  Inside the
future domain of dependence, however, ``transient solutions'' are manifestly independent of
boundary conditions.]  
Depending on the evolution equations and
background spacetime used, these transients can potentially grow significantly (i.e. by a
factor of more than $10^{16}$ in amplitude).   
Under these conditions, even roundoff-level errors in initial data should produce transients
that amplify to unit 
magnitude.   Once errors reach unit magnitude, then guided by the KST results discussed above,
we expect 
 nonlinear terms
in the equations to 
generically cause these errors to grow  even more rapidly, followed shortly thereafter by 
complete failure of a numerical simulation.
In other words, if the formulation and background spacetime permit transients to amplify by
$10^{16}$, we expect numerical simulations of these spacetimes to quickly fail.

In this paper we  develop conditions which tell us when such dramatic
amplification is \emph{assured}.  
Specifically, we describe how to compute the amplification of certain transients
for  a broad class of
partial differential equations (first-order symmetric hyperbolic PDEs) that includes many
formulations of Einstein's equations.
If this amplification is larger than
 $10^{16}$, then
we know we should not evolve this formulation numerically.

\subsection{Outline of remainder of paper}
In this paper, we  analyze the growth of transients.  [Remember, in this
paper a 
transient is any solution defined in the future domain of dependence of the initial data
slice.]   Rather than study all possible 
formulations, we limit attention to a class of partial differential equations we can analyze in
a coherent, systematic fashion: first-order symmetric hyperbolic systems.  Furthermore,
because we concern ourselves only with stability and the growth of small errors, we
 limit attention to linear perturbations upon some background.  Finally, to be able
to produce concrete predictions, we restrict attention to those transients which satisfy the
geometric optics approximation.

In Sec. \ref{sec:rays} we introduce  an explicit
ray-optics-limit solution to  first-order  symmetric hyperbolic linear systems
 -- a class which includes, among its other elements, linearizations of certain formulations of
 Einstein's equations. 
We provide explicit ODEs which determine the path (i.e. ray) and amplitude of a
geometric-optics solution, in terms of initial data at the starting point of the ray. 
Then, in Sec. \ref{sec:packets}, we introduce wave packets as solutions which are confined to a
small neighborhood of a particular ray.  We further define 
two special classes of wave packet -- coherent wave packets and
prototyptical coherent wave packetes -- which, because of their
simple, special structure,  are much easier to analyze.
Finally, in Sec. \ref{sec:energy}, we  introduce and
discuss the technique (energy norms) we will use to characterize the
amplitude of wave packets.  In particular, we provide an explicit
expression [Eq. (\ref{eq:rate:prototypical})] for the growth rate of
energy of a prototypical coherent wave packet.

To demonstrate explicitly how the techniques of the previous sections can be applied to produce the growth
rate of transients, in
Sections \ref{sec:KSTtests:Rindler} and \ref{sec:KSTtests:PG} (as well as Appendix
\ref{ap:KST}) we describe by way of example how our methods can broadly be applied to the
two-parameter formulations that Kidder, Scheel, and Teukolsky (KST) have proposed \cite{KST}.
Specifically, Sections \ref{sec:KSTtests:Rindler} and \ref{sec:KSTtests:PG} will respectively
describe wave packets on flat-space (written in Rindler coordinates) and radially propagating
transients on a Schwarzchild-black-hole background (expressed in Painleve-Gullstrand coordinates).

Finally, to demonstrate explicitly how expressions for the growth rate of transients can be
used to filter out particularly pathological formulations, in Sections
\ref{sec:KSTnumerics:rindler} and \ref{sec:KSTnumerics:pg} we
use the results for the growth rates of transients obtained in 
Sections \ref{sec:KSTtests:Rindler} and \ref{sec:KSTtests:PG} to determine what pairs of 
KST parameters ($\gamma$ and $\hat{z}$)  \emph{guarantee} significant 
amplification of some transient propagating on a Rindler and Painleve-Gullstrand background,
respectively.

\subsubsection*{Guide to the reader}
While the fundamental ideas behind this paper -- the study of wave packets and the use of their
growth rates to discover ill-behaved formulations -- remains simple, when we attempted to perform
practical, accurate computations, we quickly found the simplicity of this idea masked behind 
large amounts of novel (but necessary) notation.   We therefore found it difficult to
simultaneously satisfy the casual reader --  who wants only a summary of the essential
results, and who is still evaluating whether the results and the methods used to obtain them
are worthy of further attention -- 
and the  critical reader -- who needs comprehensive understanding of our methods in
order to evaluate, duplicate, and (potentially) extend them.
We have chosen to slant the paper towards the towards the critical reader; thus this paper 
is a \emph{comprehensive} and \emph{pedagogical} introduction to our techniques.

While this paper can be consumed in a single reading, 
for the reader interested in a brief summary of the essential ideas and results, or for anyone
making a first reading of this paper, the author
recommends only reading the most essential details.  First and foremost, the reader should
understand the scope and significance of this paper (i.e. read the abstract and Sec. I).  Next,
the reader should
follow the general description of the techiques in Sections \ref{sec:rays}, 
\ref{sec:packets}, and \ref{sec:energy} in detail.    Subsequently, the reader should examine our 
demonstration that our techniques indeed give correct results for growth rates
 (cf.  the introduction to Section
\ref{sec:KSTtests:Rindler} and the summary that section's results in Section
\ref{sec:sub:rindlerNumbers:Validity}).  Finally, to understand how these techniques can be used
to discrover ill-behaved formulations, the reader should examine Sections
\ref{sec:KSTnumerics:rindler} and \ref{sec:KSTnumerics:pg}.  

The more critical reader may wish to test and verify our computations.  This reader should then
review Sections  \ref{sec:rays}, 
\ref{sec:packets}, and \ref{sec:energy} again, then  work through  Sections
\ref{sec:KSTtests:Rindler} and \ref{sec:KSTtests:PG} in detail (returning to the earlier
sections for reference as necessary).  This reader will also
benefit from the general approach to KST 2-parameter formulations  discussed in Appendix \ref{ap:KST}.

Finally, the most skeptical readers will want to examine the conceptual underpinnings of and
justifications for our every computation.  This reader should simply follow the text as
presented, but  carefully read every footnote and appendix as they are mentioned in the text.
In particular, this reader will want to review our  Appendicies \ref{ap:ValidityOfRayOptics} (for a
justification of our ray-optics techniques) and \ref{ap:identities} (for many useful identities
used in the previous appendix and elsewhere in the paper) as well as Appendix
\ref{ap:prototypesAsGeneric} (for a more detailed discussion of prototypical
coherent wave packets, a key element in our computational method).

\subsection{Connection with prior work}

\subsubsection{Study a short-time, rather than long time, instability mechanism}
First and foremost, we should emphasize that our work differs substantially from all previoius
work on this subject: we very explicitly restrict attention to   amplification over only a
short time (i.e. a light-crossing time).   On the one hand, unlike other work, because of this
restriction, our claims -- being independent of boundary conditions -- apply to \emph{all}
boundary conditions. 
On the other,  because we forbid ourselves from studying our solutions
outside the future domain 
of dependence of the initial data slice -- even though, in practice, we could draw some
elementary conclusions%
\footnote{In fact, because these solutions are high-frequency solutions, we can quite easily
  determine their interaction with most boundary conditions.  For example,
  maximally-dissipative boundary conditions (i.e. the time derivatives of all ingoing characteristic fields are set to
  zero) imply, in the geometric-optics limit, that all solutions on ingoing rays will be zero.
  In particular, that implies that, when wave packets reach the boundary, they leave without
  reflecting.  Other boundary conditions may also be easily analyzed.
} --
in this paper  we choose not to make any claims about how
a formulation of Einstein's equations will behave at  late times (i.e. its late-time stability
properties).   

\subsubsection{Study an instability mechanism, not necessarily the dominant one}
In other papers which attempt to address the stability properties of various formulations of
Einstein's equations -- for example, Lindblom and Scheel (LS) \cite{LSenNorms} -- the authors try to 
(somewhat naturally) an
understanding of the \emph{dominant} instability mechanism. 
Unfortunately,  we do not fully understand all the dominant instability mechanisms which can
occur 
in  generic combinations of 
evolution equations, boundary conditions, and background spacetimes.  Indeed,  while some
theoretical progress has been made towards estimating the dominant instability mechanisms
(i.e. LS), for generic ``reasonable'' formulations (i.e. those which we have not excluded
based on other known pathologies, such as being weakly hyperbolic), we currently can only
reliably determine how effective simulations will be by running those simulations.  
And simulations are slow.

In this paper, instead of studying the \emph{dominant} instability, we study \emph{an}
instability 
(transients) which we  can easily understand and rigorously describe.  We use this instability
to discover formulations which are known to be troublesome (i.e. which have trouble with
transients).

\subsubsection{Short-wavelength approximations}
This paper makes extensive use of geometric optics, a special class of short-wavelength
approximation.  Several authors have applied short wavelength techniques to study  the
stability of various formulations of Einstein's 
equations \cite{paADM}, \cite{paShinkaiReview}, \cite{paFR}.  These techniques, however, have
generally been applied  to systems whose coefficients do not vary in space, limiting their
validity either to very small neighborhoods of generic spacetimes, or to flat space.   Previous
analyzes have thus obtained only a description of local plane wave propagation: in
other words,  local dispersion relations.   
In this paper, with the geometric optics approximation, we describe how to glue
these local solutions together.   Such gluing is essential if we are to obtain a good
approximation to a global solution of the PDE and hence a  concrete,
reliable estimate of the amplification of a transient.  In this sense, the present paper is the
logical extension of work by Shinkai and Yoneda (see, e.g., \cite{paShinkaiReview}), an attempt at
converting their analysis to precise, specific conditions one can impose which insure that
transients do not amplify.

\subsubsection{Energy norms}
This paper also employs the energy-norm techniques introduced by Lindblom and Scheel
(LS) \cite{LSenNorms}.  Energy norms provide a completely generic approach to determining the
growth rate given a known solution and, moreover, can be used to \emph{bound} the growth of
generic  solutions.  While LS choose to apply these techniques to study a different class of
solution -- large-scale solutions whose growth  presently limits their numerical simulations --
these techniques remain generally applicable.  We use them to characterize the growth of wave packets.

\section{\label{sec:rays}Ray optics limit of  first-order symmetric hyperbolic systems}
In classical electromagnetism, certain short-wavelength solutions to Maxwell's equations can
be approximated  
by  a set of ordinary differential equations for
independently-propagating rays: a set of equations for
the path a ray follows, and a set of equations which determine how the solution evolves along a
given ray \cite{BornWolf}.  This limit is known as the ray optics (or geometric-optics) limit.
In this section, we   construct an analogous limit for arbitrary first-order symmetric hyperbolic
linear  systems.

\subsection{\label{sec:sub:defs}Definitions}
We study a specific region of 4-dimensional coordinate space
($t,\vec{x}$), on which at each point we have a $N$-dimensional (real) vector space $V$ of
``fields'' 
$u\in V$.  

\emph{Inner products}:
On the space of fields, an \emph{inner product} is a map from two vectors $u,v$ to a real
number with certain properties (bilinear, symmetric, and positive-definite).  The inner
product is assumed to be  smooth relative to the underlying 4-manifold.  
The canonical inner product
on $R^N$ (i.e. the $N$-dimensional dot-product, relative to some basis of fields which is defined
everywhere throughout space) is denoted $(,)$, and does not vary with space.  We can represent
any other inner product in terms of the canonical inner product and a map $S: V\rightarrow V$
as $(u,S v)$, where  $(u,Sv)=(Su,v)$. 

An operator $Q$ is said to be \emph{symmetric} relative to the inner product generated by $S$ if
$(u,S Qv)=(Qu,Sv)$ for all $u,v$.   
In other words, an operator $Q$ is symmetric if it is equal
to its own conjugate relative to $S$, denoted $Q^\dag$ and defined by $(u, S Q v) = (Q^\dag u,
S v)$ for all $u, 
v$.  Equivalently, the conjugate $Q^\dag$ relative to $S$ may be defined in terms of the
transpose $Q^T$ (i.e. the conjugate relative to $S=1$):
\begin{eqnarray}
\label{eq:def:conjugate}
Q^\dag \equiv S^{-1} Q^T S  \; .
\end{eqnarray}

\emph{First-order symmetric hyperbolic linear  systems (FOSHLS)}:
A first-order  symmetric hyperbolic linear  system has the form
\begin{equation}
\label{eq:def:foshl}
\left[\partial_t + A^a(x,t) \partial_a - F(x,t)\right]u(\vec{x},t) = 0
\end{equation}
for $u(x,t)$ a smooth function from the underlying 4-manifold into the $N$-dimensional space of
fields, for $A^a$ and $F$ some (generally space and time dependent\footnote{%
As a practical matter, we will limit attention in this paper to $A^a$ and $F$
varying slowly (or not at all) in time; therefore,  all time dependence in the operators
$A^a$, $F$, and $S$ may usually be neglected.  For completeness, however, we retain time
dependence for readers who may wish to apply these techniques to more generic systems.
}) linear operators on that
space, and for $A^a$ a symmetric operator relative to some 
inner product.  

If more than one inner product makes $A$ symmetric, henceforth, when talking about a specific
FOSHLS, we shall fix one
specific (arbitrary) inner product throughout the discussion, and therefore some specific $S$.

\emph{Characteristic fields and speeds}:
For all 3-vectors $p_a$,
$A^a p_a$ is symmetric relative to the inner product generated by $S$.  It has a set of
eigenvalues, eigenspaces, and (for each eigenspace) basis eigenvectors, denoted as follows:
\begin{itemize}
\item $\omega_j(t,\vec{x},\vec{p})$ are  the eigenvalues of $A^a p_a$;
\item
$B_j(t,x,p)$, where $j$ runs from 1 to the number of distinct eigenvalues of $A^a p_a$, are the
eigenspaces of $A^a p_a$; and
\item
$v_{j,\alpha}(t,\vec{x},\vec{p})$ are some orthonormal basis of eigenvectors for the space
  $B_j(t,x,p)$, where $\alpha$ runs from 1 to the dimension of $B_j$.
\end{itemize}
Because $A^a p_a$ is symmetric relative to the inner product induced by $S$, the eigenspaces are
orthogonal relative to the inner product, and the eigenspaces are complete.  Finally, at each
point $(x,p)$ and for each  eigenspace, there is a 
unique projection operator $P_j(t,x,p)$ which satisfies $P_j v = v$ if $v\in B_j$, $P_j v = 0$ if
$v\in B_k$  with $k\ne j$.

We require $A^a p_a$ and its eigenvalues, eigenspaces, and projection operators to vary
smoothly over all $x^a$ and $p_b$ in the domain.  [We do not demand the eigenvectors themselves to
be smooth save in the neighborhood of each point $(x^a,p_b)$: topological
constraints may prevent one from defining an eigenvector everywhere (i.e. for all $p_a$ given
$x^a$)
\footnote{For example, in the first-order representation of the scalar wave equation, two of
  the eigenvectors at each point $(x,p)$ are essentially  vectors transverse to the surface
  $|p|$. These cannot be extended over the sphere.
}.]

\emph{Group velocity and acceleration}:
We  define the  group velocity
 $V_j^a$
 and group acceleration $a_{j,a}$ 
via
\begin{eqnarray}
\label{eq:def:groupV}
V_j^a(t,\vec{x},\vec{p})&\equiv& \frac{\partial}{\partial p_a} \omega_j(t,\vec{x},\vec{p})  \; ,
\\ a_{j,a}(t,\vec{x}, \vec{p})&\equiv& -\frac{\partial}{\partial x^a} \omega_j(t,\vec{x},\vec{p})
\; .
\end{eqnarray}
We shall make frequent use of an alternative expression for the  group velocity, 
Eq. (\ref{eq:tool:groupV}), which is discussed in   Appendix \ref{ap:identities}.  Among other
things, Eq. (\ref{eq:tool:groupV}) implies 
\[
\omega_j(x,p) = V^a_j p_a \; .
\]

\subsection{Form of ray-optics solution}
We now construct a solution which approximately satisfies
Eq. (\ref{eq:def:foshl}).   Our method works by constructing a set of characteristics
(i.e. rays), then 
integrating some amplitude equations along each characteristic (as an ODE) to find the
amplitudes farther along the ray.  

In this section, we only introduce the results of our analysis.  In  Appendix
\ref{ap:ValidityOfRayOptics}, we provide a more comprehensive justification of our ray-optics
approach.

\subsubsection*{Ray-optics solution}
Rather than express our solution in terms of the original $N$-dimensional variable $u$, we
introduce $N+1$ new 
variables $d_{j,\alpha}$ and $\phi$ and parametrize the original state by
\begin{equation}
\label{eq:defImplicit:ub}
u = \bar{u} e^{i \phi }
\end{equation}
where we further expand $\bar{u}$ in terms of the eigenvectors $v_{l,\alpha}$ of $A^a \partial_a \phi$
at each point $(t,\vec{x})$: 
\begin{equation}
\label{eq:defImplicit:ds}
\bar{u} = \sum_l \sum_\alpha d_{l,\alpha}(t,x,\partial \phi) 
 v_{l,\alpha}\left(t,x, \partial
  \phi\right) \; .
\end{equation}
[For notational clarity, the arguments $t$, $\vec{x}$, and $\partial_a \phi$ to the functions
$\phi$, $v_{l,\alpha}$, and $d_{l\alpha}$ will in the following be usually omitted.]

In terms of these new variables, a ray-optics solution is a solution to the following
equations, for some fixed $j$: 
\begin{subequations}
\label{eq:trial:Net}
\begin{eqnarray}
\label{eq:trial:Phase}
0&=&\left[\partial_t + V_j^a(x,\partial \phi) \partial_a \right] \phi \; ,  \\
\label{eq:trial:PolarizationOffdiag}
0&=&d_{l,\beta}  \quad \text{for} \; l\ne j  \; , \text{ and}\\
\label{eq:trial:PolarizationConfusing}
0 &=& \left[\partial_t
   +  V_j^a \partial_a \right] d_{j,\alpha}
   \\
  & &  + \sum_\beta d_{j,\beta}\bigg( v_{j,\alpha},\; 
  S(\partial_t + A^a \partial_a -F)v_{j,\beta} \bigg) \; . \nonumber  
\end{eqnarray}
\end{subequations}
When we substitute solutions to the ray-optics equations [Eq. (\ref{eq:trial:Net})] back
into the original FOSHLS [Eq. (\ref{eq:def:foshl})], as described in detail in 
Appendix \ref{ap:ValidityOfRayOptics}, we find 
the geometric-optics solutions are
excellent approximate solutions to the original PDE, so long as certain mild conditions
continue to hold [e.g. the oscillations in $\phi$ remain rapid compared to any other length or
time scale].

\subsection{Interpreting the geometric-optics equations}
We introduce the geometric-optics solution precisely because it simplifies the PDE -- in
particular, because it converts the problem of solving a general  PDE
[Eq. (\ref{eq:def:foshl})] into the problem of solving uncoupled ODEs
[Eq. (\ref{eq:trial:Net})]. 
Specifically, these ODEs consist of the the phase equation [Eq. (\ref{eq:trial:Phase})] --
which determines the path of the ray
leaving a 
point $\vec{x}$ consistent with initial data for $\phi$ with gradient $\partial_a
\phi(\vec{x})$ -- and the polarization  equations
[Eqs. (\ref{eq:trial:PolarizationOffdiag}) and (\ref{eq:trial:PolarizationConfusing})]  --
which  allow us to propagate the $d_{l,\alpha}$ along each ray. 

But while these equations are now ODEs, their structure is not particularly transparent.  In
this section, we rewrite the phase equation [Eq. (\ref{eq:trial:Phase})] and the polarization
equation [Eq. (\ref{eq:trial:PolarizationConfusing})] to better emphasize their properties and
physical interpretation.

\subsubsection{Path of the ray}
The physical significance of the phase equation [Eq. (\ref{eq:trial:Phase})] becomes much
easier to appreciate when it is
rewritten in first-order form.   When we differentiate that expression and re-express the
result as an equation for $k_a\equiv \partial_a \phi$, we find
\begin{eqnarray}
0&=&\partial_t k_a  + V_j^b \partial_b k_a + [\partial_a V_j^b(x,k)] k_b   \nonumber \\
\label{eq:trial:PhaseRayForm}
 &=& \partial_t k_a + V_j^b \partial_b k_a - a_{j,a}(x,k)  \; .
\end{eqnarray}
[While $k$ does depend on $x$, because $(\partial_{k_c} V_j^b) k_b =0$ the last term in the
first line does indeed simplify into $-a_{j,a}$, as stated].  Solutions to this PDE may be
constructed by gluing together  solutions to 
the following pair of coupled ODEs for $\vec{x}(t)$ and $\vec{k}(t)$:
\begin{subequations}
\label{eq:pathNet}
\begin{eqnarray}
\label{eq:path1}
\frac{d x^a}{dt} &=& V^a_j(\vec{x},\vec{k}) \\
\label{eq:path2}
\frac{d k_a}{dt} &=& a_{a,j}(\vec{x},\vec{k})  \; .
\end{eqnarray}
\end{subequations}
By using the definitions of $V^a_j$ and $a_{a,j}$, we find these are precisely Hamilton's
equations, using $\omega_j(t,x,k)$ as the Hamiltonian.

These two equations define the rays (i.e. characteristics).
Given initial data for $k_a$ which has  $k_a=\partial_a \phi$ in a 3-dimensional
neighborhood of a point,
we have a unique ray emanating from each point in that neighborhood.  Solutions to Eq. (\ref{eq:trial:PhaseRayForm}) follow from joining the resulting rays emanating from each point in the neighborhood
together; and solutions for $\phi$ [i.e.  Eq. (\ref{eq:trial:Phase})] follow by integrating the phase out along each ray.

\subsubsection{\label{sec:sub:polarizeSimplify}Propagating polarization along ray}
In practice, the polarization equation [Eq. (\ref{eq:trial:PolarizationConfusing})] is
difficult to interpret: since it 
involves spatial derivatives of basis vectors, and since we have freedom to choose our basis
vectors  $v_{j,\alpha}$ arbitrarily within each subspace $B_j$, we cannot transparently disentangle
meaningful terms from convention-induced effects.  

To constrain the basis and simplify the
equation, we sometimes choose a basis in the neighborhood of the ray of interest which
satisfies the \emph{no-rotation condition} [discussed at greater length in Appendix \ref{ap:sub:norotate}]:
\begin{equation}
\label{eq:trial:PolarizationBasisPropagate}
\left(v_{j,[{\alpha}}, S(\partial_t + A^a \partial_a) v_{j,{\beta}]}   \right) = 0 \; .
\end{equation}
where the square brackets denote antisymmetrization over $\alpha$ and $\beta$ [i.e. $X_{[\alpha
  \beta]} = (X_{\alpha \beta} - X_{\beta\alpha})/2$].   
The no-rotation condition completely constrains the antisymmetric part of an operator [i.e. the left side 
of Eq. (\ref{eq:trial:PolarizationBasisPropagate})]; the condition that the basis vectors
$v_{j,\alpha}$
remain orthogonal constrains that operator's symmetric part; and therefore the basis
$v_{j,\alpha}$ is   necessarily completely specified at any point along a ray in terms of
initial data for the basis.

Using the no-rotation condition, 
we find the polarization equation becomes the less-arbitrary expression
[Appendix \ref{ap:sub:reorg}]:
\begin{eqnarray}
\label{eq:trial:PolarizationEasy}
0 &=& \left(\partial_t + V_{j}^a \partial_a + \frac{1}{2} \partial_a V^a_j(x,\partial\phi)
   \right) d_{j,\alpha}  \\
 & &  - \sum_\beta  \left(v_{j, \alpha},  
    \left[SF +  \frac{1}{2}\partial_t S+ \frac{1}{2}\partial_a (S A^a)\right]v_{j,\beta}  \right) d_{j,\beta}  \nonumber
\end{eqnarray}
where in the above $v_{j,\alpha}$ is a no-rotation basis.
In Sec. \ref{sec:packets} we will use this expression to motivate the definition of
prototypical coherent wave packets, which have an exceedingly simple growth rate.

\subsection{When do geometric-optics solutions exist?}
Given initial data (say, for $k_a$ and $d_{j,\alpha}$ on some initial compact region), we can
in practice always find a solution to the geometric-optics equations
[Eq. (\ref{eq:trial:Net})] valid for some small interval $\delta t$   (i.e. by using general
PDE existence theorems, like Cauchy-Kowaleski).    However, for general initial data we cannot
solve the phase equation [Eq. (\ref{eq:trial:Phase})] for an arbitrary time  $T$.
By way of example, even if we find  each individual ray [i.e. each  solution to
Eq. (\ref{eq:pathNet}) emanating from each initial-data point] emanating from our initial data
region out to time $T$, these rays may 
cross before time $T$, 
rendering the geometric-optics solution for $d_{j,\alpha}$ both singular and inconsistent at
the ray-crossing point. 
(A similar problem arises in classical geometric optics.)
Furthermore, depending on the structure of $A^a$, certain rays may not even admit extension to
time $T$ (i.e. certain rays may be be future-inextendible, precisely like rays striking
singularities in GR). 

A proper treatment of these technical complications is considerably beyond the
scope of this paper.  In practice, we will assume we have chosen initial data so that our
  geometric-optics solution 
can be evolved to any time $T$, unless it involves transport into a
manifest singularity (i.e. a singularity of the spacetime used to generate the FOSHLS) before
time $T$.
Furthermore, we will assume the solution is \emph{well-behaved} -- that is, the congruence
has  finite values for $k_a$,  $V^a_j$, $a_{a,j}$ and their first derivatives.  With a
well-behaved solution to the phase equation [Eq. (\ref{eq:trial:Phase})], we may always find a
finite, consistent 
solution to the polarization  equation [Eq. (\ref{eq:trial:PolarizationConfusing})] in terms
of the initial data.\footnote{%
Since the polarization equation [expressed as Eq. (\ref{eq:trial:PolarizationConfusing}) or as
Eq. (\ref{eq:trial:PolarizationEasy})] is linear in the polarization fields $d_{j\alpha}$, it
therefore admits well-behaved solutions for the evolution of $d_{j,\alpha}$ along a
well-behaved ray so long as the linear operators present in that equation are well-behaved.
}

\section{\label{sec:packets}Defining wave packets}
In Sec. \ref{sec:rays}, 
we have constructed approximate solutions to linearized first-order symmetric hyperbolic
PDEs in the geometric optics limit.  These solutions are constructed by integrating ODEs for
(and along) rays [Eq. \ref{eq:trial:Net}].  Since
 each ray evolves independently, we are naturally led to consider \emph{wave
packets} -- that is, ray-optics solutions which are nonzero only in a (4-d) neighborhood of
some (4-space) ray.

In this section, we outline how wave packets may be generally constructed.  We also describe
the two special classes of wave packets, coherent wave packets and prototypical coherent wave
packets (PCWP), which will be the focus of discussion henceforth.

\subsection{Constructing  wave packets}
A wave packet that persists for a time $T$ is some solution to the geometric-optics equations
[Eq. (\ref{eq:trial:Net})]
which is nonzero only in some small neighborhood of a ray (i.e. nonzero only within some
coordinate length $\delta$ from the central ray).  

From a constructive
standpoint, while we can easily construct solutions from initial data for 
 $k_a$ and $d_{j\alpha}$, we have no transparent way, besides solving
the equations themselves, to determine whether a particular set of initial data for $k_a$ even generates a
congruence which exists and remains well-behaved (e.g. $\partial^a V_b$ and $\partial_a k_b$
both finite) for time $T$, let alone whether the specific combination of initial data for $k_a$
and $d_{j\alpha}$ yields a geometric-optics solution with support only within a given distance
$\delta$ from a ray.

Still, physically we \emph{expect} we can avoid these technical complications.  For
example, we \emph{expect} that, for all rays of physical interest, we can extend the central
ray of interest to time $T$ (i.e. characteristics of physical interest can be extended as long
as physically necessary).  We \emph{expect} that singular congruences $k_a$  can be avoided
by proper choice of 
initial $k_a$ data  (e.g. the ray equations do not require all congruences near the ray of
interest to diverge or 
come to a focus).    
And given a well-behaved congruence, we expect we can always choose initial data for
$d_{j\alpha}$ in a 
sufficiently small neighborhood so the solution for $d_{j,\alpha}$ is nonzero only within some
fixed distance $\delta$ from the central ray.

Thus, as a proper treatment of these technical complications is considerably beyond the
scope of this paper, we shall henceforth simply assume that a wave packet solution can
always 
be constructed about any ray of physical interest.

\subsection{Specialized wave packets I: Coherent wave packets}

Since rays propagate independently, one can choose arbitrary initial data, and in particular
arbitrary  polarization
directions $w$, and still obtain a wave-packet solution.  Here, $w$ is defined by 
\begin{equation}
\label{eq:def:w}
w \equiv \bar{u}/|\bar{u}| \qquad |\bar{u}|\equiv [(\bar{u}, S \bar{u})]^{1/2} \; .
\end{equation}

We prefer to further restrict attention to those wave packets which have a single, dominant
polarization direction $w$ present initially (and therefore for all time).  In other words, we
require $w$ vary slowly across the wave packet's spatial extent.  Wave packets with this
property we denote \emph{coherent wave packets}.

\subsection{\label{sec:sub:packets:prototypical}Specialized wave packets II: Prototypical
  coherent wave packets (PCWPs)}
While coherent wave packets have a simple polarization structure, characterized by some
polarization direction $w$, this polarization structure need not necessarily have a transparent
relationship to the terms present in the polarization equation
[Eq. (\ref{eq:trial:PolarizationConfusing}); or equivalently
Eq. (\ref{eq:trial:PolarizationEasy}) if we use a no-rotation basis]. 
Therefore, we define \emph{prototypical coherent wave packets} (PCWPs) as wave packets which have at
each time their polarization direction $w$ equal to  one of the eigenvectors
$f_{j}^{(\mu)}$ of the operator $O_j$:
\begin{eqnarray}
\label{eq:def:Oj}
O_j &\equiv& P_j\left\{
  F +\frac{1}{2} S^{-1}\left[\partial_t S + \partial_a(S A^a)\right]
  \right\} P_j  \\
\label{eq:def:fj}
O_j f_j^{(\mu)} &\equiv& o_{j\mu} f_{j}^{(\mu)}
\end{eqnarray}
where  $\mu$, running from $1$ to the dimension of $B_j$,  indexes the
 eigenvectors of $O_j$. 
For simplicity, we assume $O_j$ has a complete set of eigenvectors.\footnote{The behavior of the polarization equation when $O_j$ has Jordan blocks is
  straightforward (i.e. we converge to some specific eigenvector in the Jordan block; we obtain
  no change to the final predictions for exponential growth 
  rates; we only add at most a polynomial in $t$ to the amplitude functions) but tedious to
  describe in detail.  Moreover, in all physically interesting cases we have examined, Jordan
  blocks have not    appeared in $O_j$; we have been able to choose a complete set of basis
  eigenvectors.
}

If PCWPs exist, we expect -- because of their relationship to the terms of
the polarization equation [Eq. (\ref{eq:trial:PolarizationEasy})] -- the propagation of their
polarization will be much easier to understand.   Most notably, as we will show in the next section
[Sec. \ref{sec:energy}], prototypical coherent wave packets have particularly
simple  
expressions for their growth rates [i.e. Eq. (\ref{eq:rate:prototypical})].

PCWPs will exist as exact solutions to the polarization equation 
[Eq. (\ref{eq:trial:PolarizationEasy})] only  in certain special
circumstances; for example, 
most of the polarizations to be discussed in Sections 
\ref{sec:KSTtests:Rindler} and \ref{sec:KSTtests:PG} admit exact PCWP solutions.  However, as
demonstrated in more detail in Appendix \ref{ap:prototypesAsGeneric}, 
we do not expect the polarization equation  to
generically  admit PCWP solutions.

Nonetheless, as discussed in greater detail in Appendix \ref{ap:prototypesAsGeneric}, a PCWP
with $w=f_j^{(\nu)}$ is a 
good approximate solution to the polarization equation when the eigenvalue $o_{j\nu}$ of $O_j$ is
sufficiently large.  Indeed, by rewriting the polarization equation in the basis
$f_j^{(\mu)}$, we can show \emph{generic} coherent wave packets will rapidly converge to a PCWP
with 
$w=f_s^{(\nu_o)}$ for $\nu_o$ indexing the eigenvalue of $O_j$ with largest real part.  In
other words, based on Eq. (\ref{eq:rate:prototypical}), when coherent wave packets grow
quickly, they can always be well-described by a PCWP.

\section{\label{sec:energy}Describing and bounding the growth rate of wave packets}
Since a wave packet is narrow and we care little about its precise spatial extent, we commonly
characterize the wave packet by a single number (e.g. a peak amplitude) rather than a generic
distribution of polarization over space.   Unfortunately, the maximum value of the amplitudes
$d_{j,\alpha}$ depend on the
spatial extent of the wave packet -- in other words, it depends on our choice of congruence,
rather than the central ray itself.  

Because the amplitude function is subject to  focusing effects (through the term $\partial_a
V^a$), we choose to 
describe the magnitude of the wave packet by the magnitude of its energy norm.  Introduced by
Lindblom and Scheel (LS), the energy norm
is an integral
quantity analogous to  energy   \cite{LSenNorms}; and, like the energy of a wave packet solution to
Maxwell's equations, the energy norm will not be susceptible to focusing effects.

In this section, we describe how energy norms can be used to characterize the magnitude of wave
packets.  We also obtain special expressions for the growth rates of coherent wave packets
[Eq. (\ref{eq:rate:coherent})] generally and
 prototypical coherent wave
packets [Eq. (\ref{eq:rate:prototypical})] in particular. 

Also, for completeness,  in Appendix \ref{ap:energybound} we provide an
explicit, rigorous bound for the growth rate of energy which will not be otherwise used in the
paper.

\subsection{Energy norms and the magnitude of geometric-optics solutions}
Lindblom and Scheel define the energy norm by way of two quadratic functionals of a solution
$u$ [LS Eqs. (2.3) and (2.8)].  When expressed in terms of our notation, these functionals are
terms of our notation are
\begin{eqnarray}
\label{eq:def:E}
\epsilon &\equiv & (u^*, S u) \qquad E \equiv \int \mu d^3 x \; \epsilon  \; .
\end{eqnarray}
Unlike LS, we do not generically have a preferred spatial metric; we therefore replace the
factor $\sqrt{g}$  present in LS Eq. (2.8) by the more generic $\mu$.\footnote{
Unlike LS,  we are not necessarily working with a metric space; therefore, we have
no preferred measure on the coordinate space and therefore allow for an arbitrary,
as-yet-undetermined measure factor $\mu$.
}

We may substitute in the expressions appropriate to a ray-optics solution to obtain excellent
approximations to the energy.  By way of example, the energy $E_j$ of a geometric-optics
solution propagating in the $j$th polarization  may be expressed as
\begin{eqnarray}
\label{eq:ex:Ej}
E_j &\approx& \int \mu d^3 x \;  
     \sum_{\alpha,\beta} d_{j,\alpha}^* d_{j,\beta} \left(v_{j,\alpha}^*, \; S v_{j,\beta}
     \right) \\
   &=& \int \mu d^3 x \;  \sum_\alpha |d_{j,\alpha}|^2 \nonumber
\end{eqnarray}
where the terms neglected are small in the geometric optics limit and where the second line
holds because by construction the basis $v_{j,\alpha}$ is orthonormal.

\subsection{Energy norms and the growth rate of wave packets}
Following the techniques of Lindblom and Scheel, we can use energy norms and conservation-law
techniques  to obtain a general
expression for the growth rate of a wave packet.  

To follow their program, we must generate a conservation law.  Define, therefore, 
 an energy current $j^a$ [i.e. LS Eq. (2.4)]
\begin{eqnarray*}
j^a &\equiv&  (u^*, S A^a u)  \; .
\end{eqnarray*}
The quantities $\epsilon$ and $j^a$  obey the conservation-law-form equation
\begin{eqnarray*}
\label{eq:def:conservationlaw}
\partial_t \epsilon  + \mu^{-1} \partial_a \left(\mu j^a \right) =   (u^*, S F {u}) + (F {u}^*, S{u})\\
 +   \left({u}, \left[\partial_t S +\mu^{-1}\partial_a \left(\mu S A^a\right) \right]{u}\right)
 \;  \nonumber 
\end{eqnarray*}
[i.e. the analogue of LS Eqs. (2.5) and (2.6)].

For a wave-packet solution, which is concentrated at each time to a small spatial region, the
current $j^a$ drops to zero rapidly, and is in particular zero at the manifold boundary.
As a result, when we integrate the conservation law,
we find the  energy obeys the equation
\begin{subequations}
\label{eq:dEdt}
\begin{eqnarray}
\label{eq:dEdt:Explicit}
\frac{dE}{dt}  &=&\int \mu d^3 x  \left({u}^*, S Q u \right )  \\
\label{eq:def:Q}
 Q &\equiv& F+S^{-1}\left[F^T S + \partial_t S +\mu^{-1} \partial_a (\mu S A^a)\right]
\end{eqnarray}
\end{subequations}
where $F^T$ is  defined so $(u,Fv)=(F^T u,v)$ for all $u$, $v$ (i.e. $F^T$ is the
transpose). [In LS, the analogous equations are (2.7) and (2.9); in our case, however, we have
no surface term involving $ j_a$ because the solution falls off rapidly away from the wave packet.]

We can show   $Q$ is symmetric relative to $S$.\footnote{
Because $S$ and $SA^a$ are symmetric relative to the canonical inner product, so are their
derivatives.  And if  $T$  is symmetric relative to the canonical inner product, then $S^{-1}T$
is symmetric relative to the inner product generated by $S$.
}
We can also show that  that  $Q$ is closely related to the symmetric part of the operator $O_j$
[Eq. (\ref{eq:def:Oj})]:
\begin{equation}
\label{eq:tool:QasOj}
P_j Q P_j = O_j + O_j^\dag + \frac{\partial_a \mu}{\mu} P_j V_j^a \; .
\end{equation}

\subsection{\label{sec:sub:rate:coherent}Energy norms and the growth rate of coherent wave packets}

Since coherent wave packets are both localized and possess a well-defined polarization
direction $w$, we find  Eq. (\ref{eq:dEdt}) becomes, for coherent wave packets,
\begin{eqnarray}
\label{eq:rate:coherent}
\frac{1}{E} \frac{dE}{dt} &\approx& 
\left({w}^*,  S Q {w}  \right) 
\end{eqnarray}
where the right side is evaluated at the location of the wave packet at the current instant.

Because we still need the appropriate polarization direction $w$ to make use of the above
expression --  a direction we can only obtain from the polarization equation
[Eq. (\ref{eq:trial:PolarizationEasy})] -- Eq. (\ref{eq:rate:coherent}) provides only an alternate
perspective on the growth of wave packets, not an entirely independent approach to the
evolution of the amplitude.

\subsection{\label{sec:sub:rate:prototypical}Energy norms and the growth rate of PCWPs}
In the special case of a PCWP, however, we do know the polarization direction $w$:
it is one of the normalized eigenvectors $f_j^{(\mu)}$ of the operator $O_j$ [see
Sec. \ref{sec:sub:packets:prototypical}].
In this case, we find the energy growth rate for a PCWP with
$w=f_j^{(\mu)}$ to be
\begin{equation}
\label{eq:rate:prototypical}
\frac{1}{E_{j\mu}}\frac{dE_{j\mu}}{dt} 
  = o_{j\mu} + o_{j\mu}^* + \frac{\partial_a \mu}{\mu} V_j^a \; .
\end{equation}
[Here, we used  Eq. (\ref{eq:tool:QasOj}) in Eq. (\ref{eq:rate:coherent}).]

\section{\label{sec:KSTtests:Rindler}Geometric optics limit of KST:
  Rindler}

In the previous sections (Sections \ref{sec:rays}, \ref{sec:packets}, and \ref{sec:energy}), we
have developed a procedure for computing the evolution and amplification of ray-optics
solutions in  general and prototypical coherent wave packet solutions in particular.   To
provide a specific  demonstration of these methods, we demonstrate how to construct the
geometric optics limit (as described in 
Section \ref{sec:rays}) and compute the growth rate of wave packets (as described in Sections
\ref{sec:packets} and \ref{sec:energy}) when the first-order hyperbolic system is the
2-parameter first-order 
symmetric hyperbolic system Kidder, Scheel, and Teukolsky introduced (see their Section II J),
linearized about a flat-space background in Rindler coordinates.

Our computations in this section proceed as follows.
First, we review Rindler coordinates and the effects of using Rindler coordinates as the
background in the linearized KST equations.  We then describe the limited set of rays we will
study (i.e. rays that propagate only in the $x$ direction).
Subsequently, we construct the explicit form of the polarization equation
 [Eq. (\ref{eq:trial:PolarizationConfusing})] for packets that propagate only in $x$.  [The
 analysis simplifies substantially because the basis vectors used do not vary with $x$;
 therefore, the derivatives present in Eq. (\ref{eq:trial:PolarizationConfusing}) disappear.]
The analysis of the polarization equation leads us directly to an an explicit expression for
the growth of energy of a coherent wave packet [Eq. (\ref{eq:dEdt})] in general and a
prototypical coherent wave packet in particular [Eq. (\ref{eq:rate:prototypical})].

Finally, to verify our expressions  give an accurate description of the growth of
PCWPs, we compare them them against the results of numerical
simulations.

\subsection{Generating the FOSHLS using the background Rindler space}
Flat space in Rindler coordinates is characterized by the metric
\begin{equation}
ds^2 = - x^2 dt^2 + (dx^2 +dy^2 +dz^2) \; ,
\end{equation}
for $x>0$.  Using this spacetime as a background, we can linearize the KST 2-parameter
formulation to generate a  FOSHLS of the form of Eq. (\ref{eq:def:foshl}) -- and in
particular find
 explicit forms for the operators $A^a$ and $F$.  For example, we find that the principal part
 has the form [KST Eq. (2.59), along with the definition of 
$\hat{\partial}_o$ in KST  Eq. (2.10)]:
\begin{subequations}
\label{eq:ppRind}
\begin{eqnarray}
\label{eq:ppRind:1}
\partial_t  \delta g_{ij}  &\simeq& 0 \\
\label{eq:ppRind:2}
\partial_t \delta P_{ij} + x g^{ab} \partial_a \delta M_{bij} &\simeq& 0 \\
\label{eq:ppRind:3}
\partial_t  \delta M_{kij} + x \partial_k \delta P_{ij} &\simeq& 0 
\end{eqnarray}
\end{subequations}
As the right-hand sides of these equations are very long, we shall not provide them, or an explicit form
for $F$, in this paper.

Using the FOSHLS obtained by linearizing, we can proceed generally with any linear analysis,
including a construction of the geometric-optics limit.

\subsection{\label{sec:sub:rindler:LocalPlaneWaves}Describing local plane waves by diagonalizing $A^a \hat{x}_a$}
The geometric-optics limit is a short-wavelength limit.  Naturally, then, the first step towards
the geometric-optics limit is understanding the plane-wave  solutions in the neighborhood of a
point.  We find these solutions by 
substituting into Eq. (\ref{eq:def:foshl}) the form $ u \propto u_o \exp i( k\cdot x - \omega
t)$; assuming $k$ and $\omega$ are large, so we may disregard the right side; assuming both
$u_o$ and 
$A^a$ are locally constant; and then solving for $u_o$ and the relationship between $k_a$ and $\omega$.  In other words, we find those local-plane-wave solutions by
diagonalizing $A^a k_a$, 
as discussed in Sec. \ref{sec:rays}, to find eigenvalues $\omega_j$ and eigenvectors
$v_{j,\alpha}$, where $j$ indexes the resulting eigenvalues and $\alpha$ indexes the degenerate
eigenvectors for each $j$.

Because the principal part is both simple and independent of the two KST parameters ($\hat{z}$
and $\gamma$), we can diagonalize it by inspection.  For every propagation direction, the
eigenvalues are precisely $\omega_s(x,k)=s |k|$
for $s=\pm 1,0$.  For our purposes, we study only propagation in the $x$ direction.  Thus, we
need only the eigenfields of $A^a \hat{x}_a$, which 
are [see KST Eq. (2.61) and also Appendix \ref{sec:sub:basis}]
\begin{subequations}
\label{eq:basisRindler}
\begin{eqnarray}
\label{eq:basisRindler:1}
U^g_{ab} &=& g_{ab} \\
\label{eq:basisRindler:2}
U^{0}_{y,ab} &=& M_{yab}
    \\
\label{eq:basisRindler:3}
U^0_{z,ab} &=& M_{zab}
   \\
\label{eq:basisRindler:4}
U^\pm_{ab} &=& \frac{1}{\sqrt{2}}\left(P_{ab} \pm  M_{xab}\right)
\end{eqnarray}
\end{subequations}
These expressions may be interpreted as equivalent to the basis vectors $v_{j,\alpha}$, as
discussed in Appendix \ref{ap:KST} [see Appendix \ref{sec:sub:basis}, and in particular
Eq. (\ref{eq:basis})].

\subsection{\label{sec:sub:rindlerDerivePropagation}Deriving the polarization and energy equations, for propagation in the $x$
  direction on the light cone}
In this section, we describe how to construct and analyze the polarization equation
[Eq. (\ref{eq:trial:PolarizationConfusing})] and energy equation [Eq. (\ref{eq:dEdt})] for
wave packets propagating in the $x$ direction.   For technical convenience,  we limit attention
to rays which propagate on the light cone -- in other words, which travel on one of the 
two null curves of the metric:
\[
dx/dt = s x 
\]
for $s=\pm 1$.

\subsubsection{Essential tool: Diagonalizing $P_s F P_s$}
We have the polarization equation [Eq. (\ref{eq:trial:PolarizationConfusing})] and a basis
[Eq. (\ref{eq:basisRindler}), or equivalently Eq. (\ref{eq:basis})]; the application is
straightforward.    
We can, however, substantially simplify our expression by changing the
basis used to expand $\bar{u}$  from $v_{j,\alpha}$ to the basis of eigenvectors $f_s^{(\mu)}$ of $P_s F
P_s$, defined by the normalized solutions to
\[
 F f_{s}^{(\mu)} = \zeta_{s,\mu} f_{s}^{(\mu)} \; .
\]  
[Equivalently, we may define these eigenvectors in component fashion.  For each $s$, the matrix
$(v_{s,\alpha}, F v_{s,\beta})$ admits a complete set of normalized eigenvectors
$f_{s,\alpha}^{(\mu)}$:
\[
\sum_\beta (v_{s,\alpha}, F v_{s,\beta}) f_{s,\beta}^{(\mu)} = \zeta_{s,\mu}
f_{s,\alpha}^{(\mu)} \; .
\]  
Using these eigenvectors, we regenerate $f_{s}^{(\mu)} = \sum_\alpha
f_{s,\alpha}^{(\mu)}v_{s,\alpha}$, which are eigenvalues of $P_s F P_s$.]

These eigenvectors may be classified according to their symmetry properties under rotations
about the propagation axis $x$:
\begin{itemize}
\begin{subequations}
\label{eq:rindler:etas}
\item \emph{Symmetric-traceless-transverse 2-tensor} 
 [basis vectors correspond to the fields $U^s_{yz}$ and $(U^s_{yy} -
  U^s_{zz})/\sqrt{2}$]  One subspace corresponds to the  2-dimensional space of
  symmetric-traceless 2-dimensional tensors transverse to the propagation direction.  The
  operator $P_sFP_s$ is
  degenerate in this subspace; the single eigenvalue associated with this subspace is
  given by $\zeta_{s,t}$, defined by
\begin{equation}
\label{eq:rindler:etaTensor}
\zeta_{s,t} = -s 
\end{equation}

\item \emph{Transverse 2-vector} 
 [basis vectors correspond to the fields $U^s_{xz}$ and $U^s_{xy}$]  Another subspace corresponds to the  2-dimensional space of
  2-dimensional vectors  transverse to the propagation direction.  Again, the operator is
  degenerate on this space.  The
  eigenvalue of $F$ in this subspace is given by $\zeta_{s,v}$ for
\begin{equation}
\label{eq:rindler:etaVec}
\zeta_{s,v} = -s \frac{1+\gamma}{-1+2\gamma} 
\end{equation}

\item \emph{2-scalars} 
 [spanned by vectors corresponding to the fields $U^s_{xx}$ and $(U^s_{yy}+U^s_{zz})$]  Finally, the 2-dimensional space of
rotational 2-scalars has its degeneracy broken by $F$.  For each $s$, we find two eigenvalues,
denoted $\zeta_{s,s1}$ and $\zeta_{s,s2}$, with values
\begin{eqnarray}
\zeta_{s,s1} &=& -s \\
\zeta_{s,s2} &=& -s \frac{1+2\gamma^2}{-1+2\gamma} 
\end{eqnarray}

\end{subequations}
\end{itemize}

These eigenvectors $f$ are linearly independent.
Indeed, symmetry  guarantees that -- with the exception of the two 2-scalar eigenvectors -
most of the eigenvectors are mutually orthogonal.

\subsubsection{Polarization  equation for general geometric-optics solutions}
We can apply these eigenvectors to rewrite the polarization equation
[Eq. (\ref{eq:trial:PolarizationConfusing})] using the basis $f_s^{(\mu)}$.  Specifically, we
define $D_{j\mu}$ by the expansion $ d_{s,\alpha} = \sum_\mu D_{s\mu} f_{s,\alpha}^{(\mu)}$.
Noting our basis vectors $f_s^{(\mu)}$ are independent of space and time, we find  a set of
independent equations for the $D_{s\mu}$ of the form 
\begin{eqnarray}
\label{eq:kstresults:rindler:D}
(\partial_t + s x \partial_x ) D_{s\mu} = \zeta_{s\mu} D_{s\mu} \; .
\end{eqnarray}
This equation, along with the explicit forms for the basis vectors $f_{s}^{(\mu)}$,
tells us how to evolve arbitrary polarization initial data along our congruence.

\subsubsection{Energy equation for general geometric-optics solutions}
Similarly, we may rewrite expressions for the energy $E$ [Eq. (\ref{eq:def:E}), or
Eq. (\ref{eq:ex:Ej})] and growth rate $E^{-1} dE/dt$ [Eq. (\ref{eq:dEdt})]
using the basis  $f_{s}^{(\mu)}$.
For example, we define energy of the wave packet  by Eq. (\ref{eq:ex:Ej}), using a
measure $\mu=\sqrt{g}=1$ consistent with the flat spatial metric of the background.  We find,
using symmetry properties of the eigenvectors to simplify the sum,
\begin{eqnarray}
\label{eq:ex:rindler:energy}
 E &=& 
\int d^3 x \sum_{\mu \in\{t,v\}} |D_{s\mu}|^2  
\\ 
 &+& \int d^3 x\; 2 \text{Re} \left[ D_{s,s1}^* D_{s,s2}
  \left(f_{s}^{(s1)}{}^*, S f_{s}^{(s2)}\right) 
\right] \nonumber
\end{eqnarray}
The growth rate of energy $E^{-1}dE/dt$ can be obtained in two ways:
\begin{enumerate}
\item First, we can explicitly differentiate 
  Eq. (\ref{eq:ex:rindler:energy}), using Eq. (\ref{eq:kstresults:rindler:D}) to simplify as
  necessary. 
\item Alternatively, we can employ the general expression for the growth rate of
  geometric-optics solutions
  [Eq. (\ref{eq:rate:coherent})].    [To do so, we express $Q$ in terms of $O_s$ 
via Eq. (\ref{eq:tool:QasOj}).   Then we find the following explicit expression for $O_j$ by
using 
 Eq. (\ref{eq:tool:dSA}) from Appendix \ref{ap:KST}, which in this case tells us
\begin{equation}
\label{eq:rindler:tool:dSA}
P_s S^{-1}\left[\partial_t S + \partial_a (S A^a)\right]P_s = s P_s 
\end{equation}
when we rewrite the results of that expression in an operator, rather than component, notation.
Finally, we employ the basis $f_s^{(\mu)}$.  Because of Eq. (\ref{eq:rindler:tool:dSA}), we
know the eigenvectors $f_s^{(\mu)}$ of $P_s F P_s$ are equivalently eigenvectors of $O_s$.]
\end{enumerate} 
In either case, one concludes
\begin{eqnarray}
\label{eq:ex:rindler:dEdt}
 \frac{dE}{dt} &=& 
\int d^3 x \sum_{\mu} |D_{s\mu}|^2 [2 \text{Re}(\zeta_{s\mu})+ s ] \\
&+ & \int d^3 x \; 2 \text{Re} \left[ D_{s,s1}^* D_{s,s2}
  \left(f_{s}^{(s1)}{}^*, S f_{s}^{(s2)}\right) \right. \nonumber \\
& & \quad \times \left. (\zeta_{s,s1}^* + \zeta_{s,s2}
+ s )  
\right]
 \nonumber \; .
\end{eqnarray}
The above equations remain completely generic and apply to all ray-optics solutions that
propagate along the congruence $dx/dt=s x$.

\subsubsection{Energy equation in a special case: PCWPs}
As Eq. (\ref{eq:kstresults:rindler:D}) demonstrates, the polarizations do not change direction
as they propagate.    In other words, if a wave packet initially has only $D_{s\mu}\ne 0$ for
some specific pair of  $(s,\mu)$, then the wave packet will always have $D_{s\mu}\ne0$ only for
that $s$ and $\mu$.  Moreover, as noted in the discussion surrounding
Eq. (\ref{eq:rindler:tool:dSA}), the basis vectors $f_s^{(\mu)}$ used to define the $D_{s\mu}$
are eigenvectors of $O_s$.
Following the discussion of Sec. \ref{sec:sub:packets:prototypical}, we
call such a solution a prototypical coherent wave packet.

For a wave packet solution which is confined to the $(s\mu)$ polarization, we
need only one term in each sum to find the energy $E_{s\mu}$ and growth rate
$E_{s\mu}^{-1}dE_{s\mu}/dt$:
\begin{subequations}
\begin{eqnarray}
E_{s\mu} &= &\int d^3 x \; |D_{s\mu}|^2 \\
\label{eq:rindler:prototypicalRate}
\frac{1}{E_{s\mu}} \frac{dE_{s\mu}}{dt} &=& 2 \text{Re}(\zeta_{s\mu}) +s \; .
\end{eqnarray}
\end{subequations}
[The above expression was obtained directly from Eq. (\ref{eq:ex:rindler:dEdt}).
Equivalently, we can obtain the same result using Eq. (\ref{eq:rate:prototypical}) by way of Eq. (\ref{eq:rindler:tool:dSA}).]

To be very explicit, we find using Eq. (\ref{eq:rindler:etas}) the growth rates of the tensor
($t$) and one of the scalar ($s1$) polarizations to be constant, independent of $\gamma$ but
depending on which direction the packet propagates ($s=\pm 1$):
\begin{subequations}
\label{eq:rindler:growthrates}
\begin{eqnarray}
\frac{1}{E_{s,t}} \frac{dE_{s,t}}{dt} &=& 
\frac{1}{E_{s,s1}} \frac{dE_{s,s1}}{dt} = -s 
\end{eqnarray}
We also find the vector ($v$) and remaining scalar ($s2$) polarizations have a growth rate
which varies with $\gamma$, according to 
\begin{eqnarray}
\label{eq:rindler:growthrate:v}
\frac{1}{E_{s,v}} \frac{dE_{s,v}}{dt} &=& -s\left(2 \frac{1+\gamma}{-1+2\gamma} -1\right) \\
\label{eq:rindler:growthrate:s2}
\frac{1}{E_{s,s2}} \frac{dE_{s,s2}}{dt} &=& 
  -s \left(2 \frac{1+2\gamma^2}{-1+2\gamma} -1\right) 
\end{eqnarray}
\end{subequations}

\subsection{\label{sec:sub:rindlerNumbers:Validity}Comparing growth rate expressions to
  simulations of prototypical coherent wave pulses}
In Eq. (\ref{eq:rindler:growthrates}) we tabulated the expected growth rates of energy for each
possible coherent wave packet.  To demonstrate that these expressions are indeed correct, we
compare these predicted growth rates with the results of numerical simulations of wave packets
propagating on a Rindler background.

\subsubsection{Specific simulations we ran}
To test the validity of our expressions, we used a 1D variant of the KST pseudospectral code
kindly provided by Mark Scheel.  He developed this code to
study the linearized KST equations on a Rindler background (e.g. to produce the results shown
in Lindblom and Scheel Sec. IV A \cite{LSenNorms}). 

We ran this code at a fixed, high resolution (512 collocation points in the $x$ direction) on a
computational domain $x\in[0.01,1]$  with
various wave-packet initial data.  Specifically, we used a wave packet profile proportional to
\begin{equation}
W(x) = A \cos(2 \pi x/\lambda) \exp\left[ - (x-x_c)^2/\sigma^2 \right]
\end{equation}
with $A=10^{-5}$, $x_c = 0.55$, $\sigma = 0.1$, and $\lambda = 0.01$.   The precise
initial data used depended on the polarization we wanted:
\begin{itemize}
\item \emph{Tensor} When we wanted a tensor polarization, we used initial data for a single
  left-propagating 2-tensor component: $U^-_{xy}=W$, with all other characteristic fields zero.  In other
  words, we used initial data
$P_{yz} = M_{xyz} =W(x)/2$
with all other fields zero.
\item \emph{Vector} When we wanted a vector polarization, we used initial data for a single
  left-propagating 2-vector component: $U^-_{xz} = W$, with all other characteristic fields
  zero.  In other words, we used initial data
$P_{xz} = M_{xxz} =W/2$
with all other components zero.
\item \emph{Scalar 1} ($s1$) When we wanted to excite the left-propagating $s1$ polarization,
  we used initial data 
$P_{xx} = M_{xxx} =W/2$.
\item \emph{Scalar 2} ($s2$) After some algebra, one can demonstrate that to excite the $s2$
  polarization, we should use initial data
$P_{yy} = M_{xyy} =W/4$ and 
$P_{yy} = M_{xyy} =-W/4$.
\end{itemize}

To avoid the influences of boundaries, we only studied the results of the simulations out to a
time $t\sim 0.1$.

\subsubsection{Results}
For each polarization ($t$, $v$, $s1$, and $s2$), we found  that wave packets remained in the initial
polarization, with little contamination from other fields.  
For example, when exciting the tensor polarization, we found all fields other than  $U_{xy}$
remained small.

\begin{figure}
\includegraphics{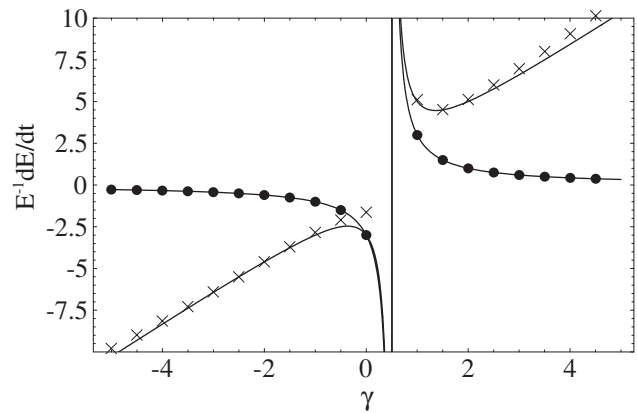}
\caption{\label{fig:rindlerRates} The two solid curves show the theoretically-predicted growth
  rates for the ``vector'' [$v$, Eq. (\ref{eq:rindler:growthrate:v})] and one of the scalar 
  [$s2$, Eq. (\ref{eq:rindler:growthrate:s2})] polarizations, when those polarizations
  propagate to the left ($s=-1$).  The circles show the results
  for numerical simulations of the vector wave packet; the crosses show the results for wave
  packets in the $s2$ polarization.  Both predictions agree very well with simulations.  
}
\end{figure}

The wave packets' energy
grew exponentially, with growth rates that agreed excellently with
Eq. (\ref{eq:rindler:growthrates}).   For example, the polarizations $s1$ and $t$ both had
growth rates consistent with unity to a part in a thousand.  Our expressions for the growth rates
 for $s2$ and $v$ also agreed well with the results of numerical simulations, as shown in
 Fig. \ref{fig:rindlerRates} for left-propagating pulses ($s=-1$).

\section{\label{sec:KSTtests:PG}Geometric optics limit of KST: PG}
In this section, we study another example of the geometric optics formalism:  the
propagation of radially-propagating 
wave  packets evolving according to
the KST 
2-parameter formulation of evolution equations, linearized about a  Painleve-Gullstrand
background. 

Our analysis follows the  same course as the Rindler case addressed in
Sec. \ref{sec:KSTtests:Rindler}.  We first 
 review Painleve-Gullstrand coordinates and the effects of using these coordinates
as the 
background in the linearized KST equations. 
Subsequently, we construct the explicit form of the polarization and energy equations
 [Eqs. (\ref{eq:trial:PolarizationEasy}) and (\ref{eq:dEdt})] for packets that propagate
 radially on the light cone.  Finally, in a departure from the Rindler pattern, we also add an
 analysis of the 
 ``zero-speed'' modes that propagate against the 
 shift vector.

\subsection{Generating the FOSHLS using a background Painleve-Gullstrand space}
A Schwarzchild hole in Painleve-Gullstrand coordinates is characterized by the metric
\begin{equation}
\label{eq:def:pg}
ds^2 = - dt^2 + \left(dr+\sqrt{\frac{2}{r}} dt\right)^2 + r^2 d\Omega^2 \; .
\end{equation}
We shall use this metric in cartesian spatial coordinates [i.e. $z=r \cos\theta$, $x=r
\sin\theta \cos \phi$, $y=r\sin\theta \sin\phi$] as the background spacetime in the KST
equations.  Linearizing about this background, we obtain the explicit FOSHLS we study in the
remainder of this section.

As before, we shall not provide the very complicated derivative-free terms (i.e. $F$)
explicitly in this paper.  The principal part, however, remains simple by design; in this case,
we have  [KST Eq. (2.59), along with the definition of 
$\hat{\partial}_o$ in KST  Eq. (2.10)]:
\begin{subequations}
\label{eq:pg:pp}
\begin{eqnarray}
(\partial_t - \beta^a \partial_a) g_{ij}  &\simeq& 0 \\
(\partial_t - \beta^a \partial_a) P_{ij} + g^{ab} \partial_a M_{bij} &\simeq& 0 \\
(\partial_t - \beta^a \partial_a) M_{kij} + \partial_k P_{ij} &\simeq& 0 
\end{eqnarray}
\end{subequations}
with $\beta^a = \sqrt{2/r}\hat{r}^a$.

\subsection{Local plane waves and diagonalizing $A^a \hat{r}_a$}
As discussed generally in Sec. \ref{sec:rays} and by way of a Rindler example in
Sec. \ref{sec:sub:rindler:LocalPlaneWaves}, to understand how wave
packets propagate radially we must first understand how local plane waves propagate radially,
which in turn 
requires we diagonalize $A^a \hat{r}_a$.   The basis vectors and eigenvalues are addressed in
detail and in a more general setting in Appendix \ref{sec:sub:basis}.  In brief, the
eigenvalues are $\omega_s(x,k) = s|k| - \beta^a k_a$ with $s=\pm 1,0$ and the eigenvectors
correspond directly to the  Rindler results
[i.e. Eq. (\ref{eq:basisRindler}), with $x\rightarrow r$; the similarity exists because we can use symmetry
without loss of generality to demand the ray propagate radially in the $x$ direction, along
$\hat{r}=\hat{x}$].

\subsection{\label{sec:sub:pgDerivePropagation}Deriving the polarization and energy equations, for radial propagation on the light
  cone}
\begin{figure}
\includegraphics{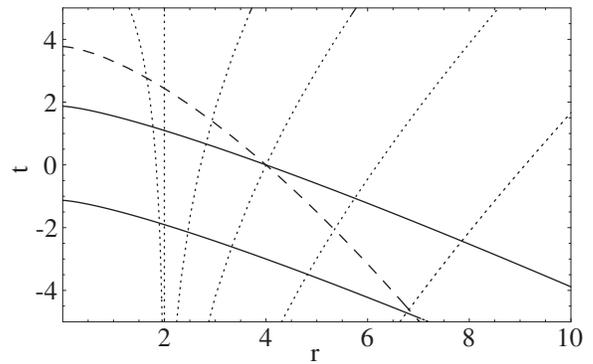}
\caption{\label{fig:pglightcones}Examples of the three types of
  radially propagating rays of the KST 2-parameter system linearized
  about a Painleve-Gullstrand background.  The solid lines show rays
  propagating inward at the speed of light ($V_r^-$).  The dotted
  lines show rays propagating ``outward'' at the speed of light
  ($V_r^+$).  Finally, the dashed curve shows the rays that propagate
  inside the light cone (at speed $V_r^0$).  The quantities $V_r^s$
  are defined in Eq. (\ref{eq:kst:pg:vg}).
}
\end{figure}

Almost half ($12$ of the $30$ characteristic fields) naturally are associated with wave packets
that propagate at the speed of light of the background spacetime (i.e. $s=\pm 1$).  In other
words, 
they propagate on characteristics that correspond to null curves of the Painleve-Gullstrand
metric [Eq. (\ref{eq:def:pg})].  For radially propagating characteristics, that means
\begin{eqnarray}
dr/dt = V^r_s \\
\label{eq:kst:pg:vg}
V^r_s \equiv s - \sqrt{2/r}
\end{eqnarray}
with $s=\pm 1$. The resulting null curve
structure is shown in Fig. \ref{fig:pglightcones}.

Because both this case and the Rindler case discussed in
Sec. \ref{sec:sub:rindlerDerivePropagation} possess rotational symmetry about the propagation
axis, the equations governing these two cases prove exceedingly similar.  The analysis follows
the same course.

\subsubsection{Essential tool: Diagonalizing $P_s F P_s$ with $s=\pm 1$}

As in the Rindler case, we will rewrite the polarization and energy equations  by using
eigenvectors $f_s^{(\mu)}$ of $P_s F 
P_s$.    Because we again have rotational symmetry about the propagation direction, we can
again decompose the eigenvectors 
into a set of two scalars ($s1$ and $s2$), a 2-vector $v$, and a symmetric-traceless-2-tensor
$t$.  The eigenvalues may be expressed using
\begin{equation}
\label{eq:result:KST:PG:etaScale}
\zeta_{s,\mu}  \equiv \bar{\zeta}_{s,\mu}/\sqrt{2}r^{3/2}
\end{equation}
where the $\bar{\zeta}_{s\mu}$ are defined by
\begin{subequations}
\label{eq:result:KST:PG:eta}
\begin{eqnarray}
\label{eq:result:KST:PG:eta1}
\bar{\zeta}_{s,s1} &=& - 3 \\
\label{eq:result:KST:PG:eta2}
\bar{\zeta}_{s,s2} &=&  \left[  
  \frac{7}{2} + 3\gamma - \frac{(33+91 \hat{z} + 24\hat{z}^2)}{4(1+3 \hat{z})(1-2\gamma)}
 \right]  \\
\label{eq:result:KST:PG:eta3}
\bar{\zeta}_{s,v} &=&  \frac{3-3\hat{z} - 5\gamma}{1-2\gamma}\\
\label{eq:result:KST:PG:eta4}
\bar{\zeta}_{s,t} &=& 1
\end{eqnarray}
\end{subequations}
The eigenspaces are, by symmetry, spanned by precisely the same fields as  in the Rindler case.
In particular, as in the Rindler case the eigenvectors do not change as we move along a ray.

\subsubsection{Polarization equation for $s=\pm 1$}
For polarizations which propagate radially on the light cone (i.e. $s=\pm 1$), the polarization
equation [Eq. (\ref{eq:trial:PolarizationConfusing})] can be written as 
\begin{eqnarray}
0 &=& \left[\partial_t
   +  V_s^r \partial_r \right] d_{s,\alpha} + s\frac{d_{s,\alpha}}{r}
   \\
  & &  - \sum_\beta d_{s,\beta}\bigg( v_{s,\alpha},\; 
  SF v_{s,\beta} \bigg) \; . \nonumber  
\end{eqnarray}
where we make use of Eqs. (\ref{eq:tool:generalPolarize1}) and
(\ref{eq:tool:generalPolarize2}) to simplify the right side, and where we observe $\partial_a
\hat{r}^a = 2/r$.  

As in the Rindler case, we may expand the amplitude $\bar{u} =\sum_\mu D_{s\mu}
f_s^{(\mu)}$ in terms of the basis $f_s^{(\mu)}$, and thereby arrive
at  polarization propagation equation precisely analogous to the Rindler result
[compare with Eq. (\ref{eq:kstresults:rindler:D})]:
\begin{equation}
\label{eq:pg:D}
\left[\partial_t + \left(s-\sqrt\frac{2}{r}\right)\partial_r\right] D_{s \mu} 
  = \left(\zeta_{s\mu}- \frac{s}{r} \right)
D_{s\mu} \; .
\end{equation}
These equations may be integrated to describe the evolution of polarization along any
individual radial ray.   

\subsubsection{Energy equation for $s=\pm 1$}
Because symmetry guarantees a close similarity between this Painleve-Gullstrand case and the
Rindler case, we find the energy $E$ of a geometric-optics-limit solution propagating on the
light cone radially inward ($s=-1$) or outward ($s=+1$) can be expressed with precisely the
same expression we used in the Rindler case: Eq. (\ref{eq:ex:rindler:energy}).  [In this case,
we again use a measure $\mu=1$ compatible with the background flat spatial cartesian-coordinate
metric.]

The rate of change of this energy, $dE/dt$, can be obtained in two ways.  On the one hand, we can directly form
$E$, convert to spherical coordinates, differentiate the resulting expression for $dE/dt$, and
use Eq. (\ref{eq:pg:D}).  On the other hand, we can  find $dE/dt$ using the general expression
of Eq. (\ref{eq:dEdt}), an expression we simplify by using 
i) the relation between $Q$ and $O_j$ given in
Eq. (\ref{eq:tool:QasOj}),
ii)
the basis $f_s^{(\mu)}$ of
eigenvectors of $P_s F P_s$,  and iii) the  expression [obtained
from Eq. (\ref{eq:tool:dSA}) and converted from a component to an operator expression]
\begin{equation}
\label{eq:pg:tool:dSA}
P_s \left[\partial_t S + \partial_a (S A^a)\right] P_s
  = -\frac{3}{\sqrt{2}r^{3/2}} P_s \; .
\end{equation}
In either case, we conclude
\begin{eqnarray}
 \frac{dE}{dt} &=& 
\int d^3 x \sum_{\mu} |D_{s\mu}|^2 
  \frac{2 \text{Re}(\bar{\zeta}_{s\mu})- 3}{\sqrt{2}r^{3/2}} \\
&+&
\int d^3 x \;  2 \text{Re} \left[ D_{s,s1}^* D_{s,s2}
  \left(f_{s}^{(s1)}{}^*, S f_{s}^{(s2)}\right) \right. \nonumber \\
& & \quad \times \left. \frac{\bar{\zeta}_{s,s1}^* +\bar{\zeta}_{s,s2} - 3}{\sqrt{2}r^{3/2}}
\right]
 \nonumber \; .
\end{eqnarray}

In particular, for  prototypical coherent wave packets -- that is, wave packets where $s$ and $\mu$ are the
   same everywhere in the packet -- we can express the growth rate of the energy $E_{s\mu}$ of
   the wave packet as
\begin{equation}
\label{eq:result:KST:PG:energyRatePolarized}
\frac{1}{E_{s\mu}}\frac{dE_{s\mu}}{dt} = 
  \frac{2 \text{Re}(\bar{\zeta}_{s\mu})- 3}{\sqrt{2}r^{3/2}} \\
\end{equation}
where $r$ is the current location of the packet.

\subsection{Deriving the polarization and energy equations, for radial propagation against the
  shift vector}
The remaining $18$ fields propagate inward against the shift vector, at speed 
$V_o = -\sqrt{2/r}$.

We shall not follow 
the same pattern we used to address propagation on the light cone 
[on a Rindler background in Sec. \ref{sec:sub:rindlerDerivePropagation} and on a
Painleve-Gullstrand background in Sec. \ref{sec:sub:pgDerivePropagation}].  In those sections,
we provided extensive discussion and background -- the explicit form of the polarization equation; a modified
form of the polarization equation in an alternative basis; explicit expressions for the growth
rate of energy general geometric-optics solutions; explicit demonstration that PCWP solutions
existed --  before finally recovering the growth rate
of PCWPs.
Instead, for pedagogical and other reasons [see Sec. \ref{sec:sub:pg:whynotcontinue}], we shall
take a briefer, more practical approach better suited to extracting precisely the information
needed to decide when some coherent wave packet can amplify a significant amount within the
future domain of dependence.

Specifically, following the arguments at the end of Sec. \ref{sec:sub:packets:prototypical}, we
expect that -- whether or not PCWPs exist as exact solutions to the polarization equation --
when the largest eigenvalue $o_{o\nu}$ of $O_o$ is  particularly large, a generic coherent
wave packet will rapidly converge to a PCWP with $w=f_o^{(\nu)}$.  In other words, we expect
that when the growth rates are large, the growth rate of generic coherent wave packets can be
obtained by finding the largest 
value of $dE/dt/E$ for PCWPs [i.e. the maximum of Eq. (\ref{eq:rate:prototypical}) over
$\mu$]. 

In short, we continue to evaluate Eq. (\ref{eq:rate:prototypical}) to get 
growth rates, though now we trust the results only when the growth rates are large.

\subsubsection{\label{sec:sub:pg:slow:quickway}Growth rate of PCWPs}

To evaluate the growth rate of  PCWPs, we must diagonalize $O_o$:
\[
O_o = P_o \left\{ F + \frac{1}{2}S^{-1}\left[\partial_t S + \partial_a (S A^a)\right]
  \right\} P_o
\]
However, from Eq.  (\ref{eq:pg:tool:dSA}) we know the term in square brackets is diagonal.
Therefore,  diagonalizing $O_o$  to obtain eigenvalues $o_{o\mu}$ and eigenvectors
$f_{j}^{(\mu)}$ is equivalent to diagonalizing $P_o F P_o$ for eigenvalues $\zeta_{o\mu}$ and
eigenvectors $f_o^{(\mu)}$ .  The eigenvalues of  
the two operators are related by
\begin{equation}
\label{eq:pg:Ojeigenvalue}
o_{o\mu} = \zeta_{o\mu} - \frac{3}{2\sqrt{2} r^{3/2}} \;
\end{equation}
We shall express the  eigenvalues $\zeta_{o\mu}$ of $P_o F P_o$ in terms of the dimensionless
rescaled quantities $L_\mu$, defined implicitly by
\begin{equation}
\label{eq:def:Lmu}
\zeta_{0\mu} = L_\mu \times \sqrt{2}/r^{3/2}
\end{equation}
Substituting 
Eq. (\ref{eq:pg:Ojeigenvalue}) into the general expression for the growth rate of PCWPs [Eq. (\ref{eq:rate:prototypical})], we find that a PCWP
in the polarization $\mu$ will have energy grow at rate
\begin{equation}
\label{eq:pg:slowrate}
\frac{1}{E_{o\mu}} \frac{dE_{o\mu}}{dt} = \left[2 \text{Re}(L_{\mu}) - \frac{3}{2} \right]
   \frac{\sqrt{2}}{r^{3/2}}
\end{equation}
where $r$ is the instantaneous location of the packet.

\subsubsection{Essential tool: Diagonalizing $P_o F P_o$ }
To obtain explicit growth rate expressions using  Eq. (\ref{eq:pg:slowrate}), we need the
eigenvalues of $P_o F Po$, expressed according to Eq. (\ref{eq:def:Lmu}). 

As in the previous two cases, the eigenspaces of $P_o F P_o$ may be decomposed into distinct
classes, depending on their symmetry properties of rotation about the propagation axis.  These
spaces are as follows:
\begin{subequations}
\label{eq:KST:pg:L}
\begin{itemize}
\item \emph{Helicity-0} a 4-dimensional space of rotational scalars (``helicity-0'' states),
  with eigenvalues given by Eq. (\ref{eq:def:Lmu}) with
\begin{eqnarray}
  L_{s1,s2} &=& \frac{-1+3\hat{z} + 18\hat{z}^2 \pm \sqrt{3} \sqrt{Y_1}}{4(1+3\hat{z})} \\
  L_{s3} &=& \frac{1}{4}(-30+19\eta +12\hat{z} - 6\eta \hat{z}) \\
  L_{s4} &=&   \frac{3}{2}(1+2\hat{z})
\end{eqnarray}
Here, we use $\eta \equiv -2/(\gamma-1/2)$ and $Y_1$ given according to an expression listed
below [Eq. (\ref{eq:def:Y1})].

\item \emph{Helicity-1} an 8-dimensional space of rotational 2-vectors (``helicity-1'' states),
  with doubly-degenerate eigenvalues given by 
\begin{eqnarray}
  L_{v1} &=& -2 \\
  L_{v2} &=& \frac{1}{2}(1+6\hat{z}) \\
  L_{v3,v4} &=& 
  \frac{3}{8}(5+8\hat{z}) +
   \frac{
      \eta(13+83\hat{z} +84 \hat{z}^2)}{32(1+3\hat{z})}\nonumber\\
  & &  \pm \frac{\sqrt{Y_2}}{32(1+3\hat{z})} 
\end{eqnarray}
Again, we use $\eta \equiv -2/(\gamma-1/2)$.
The expression for $Y_2$ is given below [Eq. (\ref{eq:def:Y2})].

\item \emph{Helicity-2} a 4-dimensional space of
 symmetric-traceless-2-tensors (``helicity-2'' states), with doubly-degenerate  eigenvalues given by
\begin{eqnarray}
  L_{t1} &=& \frac{3}{2}(1+2\hat{z}) \\
  L_{t2} &=& 2+3\hat{z}
\end{eqnarray}
\item \emph{Helicity-3} and finally a 2-dimensional space of helicity-3 states, with eigenvalue 
\begin{equation}
  L_{3} = 3(1+\hat{z}) 
\end{equation}
\end{itemize}
\end{subequations}

In the above discussion, $Y_{1,2}$ are defined by
\begin{eqnarray}
\label{eq:def:Y1}
Y_1 &=& (1+3\hat{z})(-5+5\hat{z}+24\hat{z}^2 + 36\hat{z}^3)\\
\label{eq:def:Y2}
Y_2 &=& 1296(1+3\hat{z})^2 + \eta^2(13+83\hat{z}+84\hat{z}^2)^2 \nonumber \\
  & &  - 24\eta(1+3\hat{z})(89+199\hat{z}+132\hat{z}^2)
\end{eqnarray}
and we use the shorthand $\eta \equiv -2/(\gamma-1/2)$.

\subsubsection{\label{sec:sub:pg:whynotcontinue}Aside: 
Why can't we follow the previous pattern?}
Unlike all cases previously discussed, a handful of the eigenvectors
depend weakly on position.
As a result, the use of a basis which diagonalizes $O_o$ does not offer as dramatic a
simplification as it did in our earlier analyzes of the  polarization equation
[Sections \ref{sec:sub:rindlerDerivePropagation} and \ref{sec:sub:pgDerivePropagation}].
To be explicit, if we rewrite the polarization equation in the basis $f_o^{(\mu)}$ in the
fashion of those earlier analyzes, we obtain [see Eq. (\ref{eq:tool:polarizationPrototypicalEquation})]
\begin{equation}
\label{eq:pg:slow:polarizationRewrite}
\sum_\nu D_{o\nu} M_{\mu\nu}
 =\left(\partial_t + V_o^a\partial_a 
   + \frac{1}{2} \partial_a V_o^a - o_{o\mu}\right) D_{o\mu}
\end{equation}
for $M_{\mu\nu}$ some \emph{nonzero}, position-dependent matrix coupling the various $D_{o\mu}$.

\section{\label{sec:KSTnumerics:rindler}Transients 
 and limitations on numerical simulations: Rindler}
In earlier sections, we developed -- in general 
[Sections \ref{sec:rays}, \ref{sec:packets}, and \ref{sec:energy}] 
and for specific examples [e.g., Sec. \ref{sec:KSTtests:Rindler} analyzes propagation of
transients according to the KST 2-parameter formulation of Einstein's equations, linearized
about a Rindler background]
--
tools to analyze the growth of special (i.e. prototypical coherent wave packet)
geometric-optics-limit transient solutions.
In this section, we demonstrate how these tools can be used to discover when a particular
formulation of Einstein's equations [here, some specific member of the KST 2-parameter system]
which is linearized about a specific background [here, flat space in Rindler coordinates]
admits some massively-amplified transient solution.

Specifically, in this section we apply the general tools developed in an earlier section
[Sec. \ref{sec:KSTtests:Rindler}] to determine the largest possible amplification of a
prototypical 
coherent wave packet while it remains within the future domain of dependence of some initial
data slice.  
In Sec. \ref{sec:sub:rindler:domain} we describe the initial data slice we chose and the
subset of transient solutions we studied.  In Sec. \ref{sec:sub:rindler:amplify}, we apply the
tools developed in an earlier section [Sec. \ref{sec:KSTtests:Rindler}] to determine the
amplification of each transient.  We also find an expression for the largest possible
amount a transient can amplify.  Finally, in Sec. \ref{sec:sub:rindler:restrictions}, we invert
our expression to determine which pairs of KST parameters ($\hat{z}$, $\gamma$) admit
transients that amplify in energy by more than $10^{32}$ (i.e. in amplitude by more than $10^{16}$).

\subsection{\label{sec:sub:rindler:domain}Transients studied}
\begin{figure}
\includegraphics{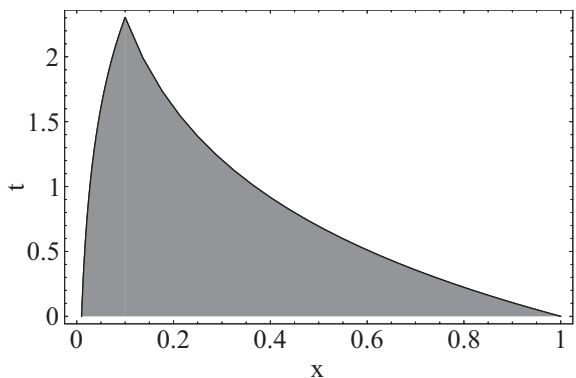}
\caption{\label{fig:rindlerDomainOfDependence}The shaded region is the future domain of
  dependence of the region 
  $x\in[0.01,1]$ for the KST 2-parameter formulation of evolution equations linearized about a
  Rindler background.  Transients are any solutions which are defined in this region.  We study
  all the prototypical coherent wave packets which propagate on the light cone (i.e. according
  to $dx/dt = \pm x$).  
}.
\end{figure}

We limit attention to the future domain of dependence of the initial-data slice $x\in[0.01,1]$
at $t=0$.  Since the KST 2-parameter formulation has fields which propagate at (but no faster than) the
speed of light, the future domain of dependence of this slice is precisely what we would obtain
using Einstein's equations: a region bounded by the two curves
$x_- \equiv 0.01 \exp t$ and $x_+ \equiv \exp(-t)$.  This region is shown in Fig. \ref{fig:rindlerDomainOfDependence}.
The future domain of dependence extends to time 
\begin{equation}
T_\text{max}\equiv \ln 10 \; ,
\end{equation} 
at which point the
two bounding curves curves intersect.

Geometric-optics solutions are defined on rays [i.e. solutions to Eq. (\ref{eq:trial:PhaseRayForm})].
While three classes of rays exist in this region -- those ingoing at the speed of light ($dx/dt =
-x$); those outgoing at the speed of light ($dx/dt = +x$); and those which have fixed
coordinate position -- we for simplicity  chose to study only the amplification of transients that
propagate on the light cone.

\subsection{\label{sec:sub:rindler:amplify}Amplification expected}
For each ray that propagates on the light cone ($dx/dt=\pm x$) within the future domain of
dependence, and for each polarization on that ray, we can compute the amplification in energy.
If $R_{s,\mu}\equiv E_{s\mu}^{-1}dE_{s\mu}/dt$ [see Eq. (\ref{eq:rindler:growthrates})], we can
express the ratio of energy of the wave packet when it exits the future domain of dependence at
time $t_\text{out}$ to the initial energy at time $t=0$ as
\[
{\cal A}_{s\mu}(x_o) 
  = E_{s\mu}(t_\text{out})/E_{s\mu}(0) = \exp \left(t_\text{out} R_{s\mu}\right) \; .
\]
We have explicit expressions for $R_{s\mu}$; we can compute $t_\text{out}(x_o,s)$ 
for each initial
point $x_o$ and for each propagation orientation (i.e. for each $s$); and we therefore can
maximize $A_{s\mu}(x_o)$ over all possible choices of initial location ($x_o$),
propagation direction ($s$), and polarization ($\mu$) to
find the largest possible ratio ${\cal A}$ of initial to final prototypical coherent wave
packet energy.

In fact, because for each polarization of prototypical coherent wave packet, the growth rates of energy is independent of time and space, the largest
amplifications possible always occur along the longest-lived rays -- in other words, along the
two bounding rays $x_+$ and $x_-$, which both extend to $t_\text{out}=T_\text{max}$.
Therefore, 
we conclude that, while within the future domain of dependence of the slice $x\in[0.01,1]$, the
largest amount the energy of any prototypical coherent wave packet can amplify is given by the
factor
\begin{equation}
\label{eq:rindler:amplify}
{\cal A} = \exp (T_\text{max} R_\text{Rind})
\end{equation}
where $R_\text{Rind}$ is given by
\begin{eqnarray}
\label{eq:num:rindlerMax}
R_\text{Rind}&\equiv& \max_{\mu,s} R_{s\mu}
   = \max_{\mu,s} \left[2 \text{Re}(\zeta_{s\mu}) + s\right]  \\
  &=&  {\max}\left(
   1, \left|2\frac{1+\gamma}{-1+2\gamma}-1\right|,
   \left| 2\frac{1+2\gamma^2}{-1+2\gamma}-1\right|
  \right) \nonumber
\end{eqnarray}

\subsubsection*{\label{sec:sub:rindler:restrictions} KST formulations which definitely possess some
  ill-behaved transient solution when linearized about Rindler}
\begin{figure}
\includegraphics{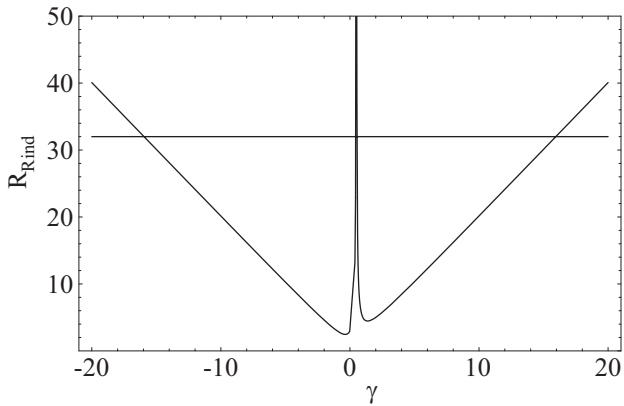}
\caption{\label{fig:rindlerRate}The solid curve is the theoretical
  prediction for the largest growth rate of wave packets that propagate on the light
  cone [Eq. (\ref{eq:num:rindlerMax})]. The 
  horizontal line is the value $32$.  According to arguments made in
  Sec. \ref{sec:sub:rindler:restrictions}, those $\gamma$ which have $R_\text{Rind}>32$ have
  some prototypical coherent wave packet which, in the future domain of dependence, amplifies
  in energy by more than $10^{32}$.
}
\end{figure}

Finally, we can invert Eq. (\ref{eq:rindler:amplify}) to find those combinations of KST
parameters ($\hat{z}$, $\gamma$) which permit some transient (in particular, some prototypical
coherent wave packet) to increase in energy by more than a factor $10^{32}$ (i.e. $10^{16}$ in
amplitude).  The condition may be expressed as either ${\cal A} > 10^{32}$ or, equivalently, as
$R_\text{Rind}>32$.  
The function $R_\text{Rind}$ is shown in Fig. \ref{fig:rindlerRate}, along with the line $R_\text{Rind}=32$.

Therefore, we know that some transient can amplify in energy by more than $10^{32}$ if
i) $\gamma>(33+\sqrt{949})/4$, ii) $\gamma<-(31+\sqrt{1077})/4$, or iii) $\gamma>29/64$ and $\gamma<(33-\sqrt{949})/4$.

\subsection{Relevance of our computation to numerical simulations}
We have demonstrated that the KST 2-parameter formulation of Einstein's equations always
admits, at any instant, prototypical coherent wave packet solutions which grow exponentially in
time.  Generically, we expect that at each instant (including in the initial data) these solutions are excited by errors in the numerical
simulation (e.g. truncation and roundoff).  They then propagate and grow; eventually, they
reach the computational boundary.  

Our calculations above  describes the largest amount any
such wave packet solution could possibly grow by the time it reaches the computational
boundary.   If that amplification factor is sufficiently large that the wave packets reach
``unit'' amplitude (i.e. whatever magnitude is needed to  couple to nonlinear terms strongly),
here conservatively assumed to be $10^{16}$, then we expect any simulation using that
particular combination of KST parameters will quickly crash.

\subsubsection*{Aside: What happens to PCWPs at late times?}
Eventually, the wave packets excited by numerical errors will reach the computational
boundary.  What happens afterward depends strongly on the precise details of the boundary
conditions.  

For example, maximally-dissipative boundary conditions (i.e. the time derivatives of all
outgoing characteristic fields are set to zero) will allow the wave packet
to leave the computational domain entirely (with some small amount of reflection that goes to
zero in the geometric-optics limit).  In this case, at late times no transient will ever
amplify 
by more than the amount described above (in Sec. \ref{sec:sub:rindler:amplify}).

On the other hand, other choices for boundary conditions could cause wave packets to reflect
back in to the computational domain.  In these circumstances, the outcomes are far more varied
-- at late times, the wave packet could potentially grow, could decay to zero, or could enter a
repetitive cycle where on average its amplitude is constant.\footnote{In fact, in this
  particular case, we expect that if a wave packet with growth rate $1/\tau$ reflects, then
  symmetry and the structure of the Rindler growth rates
  [i.e. Eq. (\ref{eq:rindler:growthrates})] 
  insures that the reflected ray has growth rate $-1/\tau$.  Therefore, on average, the wave
  packet has a zero growth rate.
}

Therefore, without some more specific proposal for boundary conditions, we cannot make useful
statements regarding the late-time development of this instability process -- or, in other
words, we cannot study
the growth of coherent wave packets for more than a light crossing time.

\section{\label{sec:KSTnumerics:pg}Transients 
and limitations on numerical simulations: PG}
In this section, we provide another example of how tools developed earlier for the analysis of
prototypical coherent wave packets --
in general 
[Sections \ref{sec:rays}, \ref{sec:packets}, and \ref{sec:energy}] 
and for specific examples [e.g., Sec. \ref{sec:KSTtests:PG} analyzes propagation of
transients according to the the KST 2-parameter formulation of Einstein's equations, linearized
about a Painleve-Gullstrand background] -- can be applied to discover which formulations of
Einstein's equations permit ill-behaved transients.

Specifically, in this section we study the propagation coherent wave packets in
the 2-parameter KST form of Einstein's evolution equations, linearized about Schwarzchild
written in Painleve-Gullstrand (PG) coordinates.  The theory needed to understand the
propagation 
and growth of radially-propagating coherent wave packets has been developed in an earlier section
[Sec. \ref{sec:KSTtests:PG}].   We apply our techniques to a handful of coherent wave packet
transient solutions, to discover conditions on the two KST parameters ($\hat{z}$, $\gamma$)
which permit  amplification of those transients' energy by a factor $1/\epsilon_e^2$ for
$\epsilon_e = 10^{-16}$.

To provide concrete examples of estimates, we assume the initial data slice contains the region
$r\in[2,10]$.  So  any influence from boundary conditions cannot muddle our computations, we
limit attention to coherent which are defined in the future domain of dependence of that slice.

\subsection{\label{sec:sub:pg:domain}Transients studied}
\begin{figure}
\includegraphics{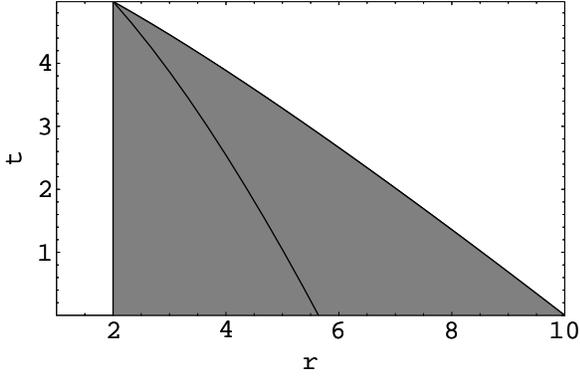}
\caption{\label{fig:pgDomainOfDependence}The shaded region is the future domain of dependence of the region
  $r\in[2,10]$ for the KST 2-parameter formulation of evolution equations.   Transients are any
  solutions which are defined in this region.  For reasons emphasized
  in the text, the rays that last for the longest coordinate time
  prove particularly helpful.
  These rays are the left and right boundaries (i.e. the horizon and a
  ray that propagates inward at the speed of light from $r=10$) and
  one ray propagating against the shift vector which emanates from
  their intersection.
}
\end{figure}

We limit attention to the future domain of dependence of the region $r\in[2,10]$ at $t=0$.
Since the KST 2-parameter formulation has fields which propagate at (but no faster than) the
speed of light, the future domain of dependence of this slice is precisely what we would obtain
using Einstein's equations: the region shown in Fig. \ref{fig:pgDomainOfDependence}. 
In particular, the future domain of dependence is bounded on the left by the generators of the
horizon (trapped at $r=2$) and on the right by rays travelling inward at the speed of light.
This ingoing ray reaches $r=2$ at the endpoint of the future domain of dependence, at time
$t=t_\text{max}$ defined by
\begin{eqnarray}
t_\text{max} &\equiv& \int_2^{10} \frac{dr}{1+\sqrt{2/r}} \\
   &=& 4 [3-\sqrt{5}+\text{csch}^{-1}(2)]
   \approx 4.98 \nonumber
\end{eqnarray}

In our future domain of dependence, we have three classes of solutions to the ray-propagation
equation [Eq. (\ref{eq:trial:PhaseRayForm})]: those ingoing at the speed of light ($V_- =
-1-\sqrt{2/r}$); those ingoing with the shift ($V_o = - \sqrt{2/r}$); and those outgoing ($V_+
= 1-\sqrt{2/r}$) [Eq. (\ref{eq:kst:pg:vg})].

\subsection{\label{sec:sub:pg:amplify}Amplification conditions}
For each of the three classes of rays ($s=\pm 1,0$) propagating radially in the future domain
of dependence [Fig. \ref{fig:pgDomainOfDependence}] and for each polarization on that ray, we
can compute the amplification in energy using $R_{s\mu} \equiv E_{s\mu}^{-1}dE_{s\mu}/dt$ [see
Eqs. (\ref{eq:result:KST:PG:energyRatePolarized}) and (\ref{eq:pg:slowrate})].  Specifically,
for a wave packet starting at $r=r_o$ at time $t=0$, propagating in the  $s$-type congruence
and in the polarization $\mu$, the energy at the time $t_\text{out}(r_o,s)$ the ray exits the
future domain of dependence is larger than the initial energy by a factor
\begin{subequations}
\begin{eqnarray}
{\cal A}_{s\mu}(r_o) &\equiv & E_{s\mu}(t_\text{out})/E_{s\mu}(0) \\
\ln {\cal A}_{s\mu}(r_o) &=& \int_0^{t_\text{out}} dt \;  R_{s\mu}  
  = \int_{r_o}^{r_\text{out}} \frac{dr}{V_r^s} R_{s\mu}
\end{eqnarray}
\end{subequations}
We then search over all $r_o$, over all propagation directions $s$, and over all polarizations
$\mu$ to find the largest amplification factor ${\cal A}$.

In fact, as in the Rindler case, we immediately know which rays produce the largest possible
amplification, so we can perform the maximization by inspection.
\begin{itemize}
\item \emph{Outgoing at light speed}: 
  Since the amplification of energy increases as $r$ gets smaller ($dE/dt/E \propto 1/r^{3/2}$)
  and with the duration of the ray in time, manifestly the generator of the horizon -- with both the longest duration and 
  the smallest $r$  of all outgoing rays --will
  provide the largest possible amplification.

Since the ray of interest has fixed radial location $r=2$, we find $\zeta_{+,\mu}$ is constant
for all polarizations.  Thus, the energy of a prototypical coherent wave packet in polarization
$\mu$ increases by a factor $A_{+\mu}$, for $A_{+\mu} = \exp(t_\text{max} \zeta_{+\mu})$.  In
other words,
\begin{eqnarray}
\ln {\cal A}_\mu
 &=& \left[2 \text{Re}(\bar{\zeta}_{+,\mu}) - 3\right] [3-\sqrt{5} + \text{csch}^{-1}(2)] \\
  &\approx & 1.245 \left[2 \text{Re}(\bar{\zeta}_{+,\mu}) - 3\right] \nonumber
\end{eqnarray}
[The values for each $\bar{\zeta}_{+\mu}$ are given in Eq. (\ref{eq:result:KST:PG:eta}).]

\item \emph{Ingoing at light speed}: The longest ray  -- namely, the right boundary of the
  future domain of dependence -- permits the greatest possible amplifications.   Thus, among
  all possible ingoing rays, the largest amplification factor for the polarization $\mu$ is
  given by ${\cal A}_{-\mu}$:
\begin{eqnarray}
\ln {\cal A}_{-\mu} 
  &=& \left[2 \text{Re}(\bar{\zeta}_{-,\mu}) - 3\right]\cdot 
      \frac{\ln 5 - 2\text{csch}^{-1}(2)}{2} \\
  &\approx& 0.323\left[2 \text{Re}(\bar{\zeta}_{-,\mu}) - 3\right] \nonumber
\end{eqnarray}
[The values for each $\bar{\zeta}_{-\mu}$ are given in Eq. (\ref{eq:result:KST:PG:eta}).]

Note that  since $\zeta_{-\mu}=\zeta_{+\mu}$, the outgoing transients trapped on the horizon grow \emph{more} 
than the ingoing ones over the same time interval\footnote{This should be expected: the ingoing and outgoing wave packets
  have similar growth rates at any given radius; we limit attention to rays which persist for a
  fixed time; and the outgoing modes we study remain closer to the horizon, where the growth
  rate is larger.
}.

\item \emph{Ingoing with lapse}: The amplification of energy increases both with ray length and
  with proximity to $r=0$ (since growth rates go as $1/r^{3/2}$).  Thus, the longest ray
  propagating at this speed contained in the  future domain of dependence gives the best
  chances.  That ray starts with $r=r_L$, with $r_L$ defined so the ray terminates at the
  horizon at $t=t_\text{max}$:
\begin{equation}
r_L \equiv \left[  
  \frac{\left(4+3 t_\text{max}\right)^2}{2}
  \right]^{1/3}
\end{equation}

  Thus, we find the largest possible amplification among those polarizations that
  have $s=0$ to be given by ${\cal A}_{0\mu}$:
\begin{eqnarray}
\ln {\cal A}_{o\mu} &=& \left[2 \text{Re}(L_\mu) -\frac{3}{2}\right]
    \times\int_{2}^{r_L} \frac{dr}{\sqrt{2/r}} 
  \frac{\sqrt{2}}{r^{3/2}} \nonumber \\
 &=&   \left[2 \text{Re}(L_\mu) -\frac{3}{2}\right] \ln \left( r_L / 2 \right)
\end{eqnarray}
[The values for each $L_\mu$ are given in Eq. (\ref{eq:KST:pg:L}).]

\end{itemize}

\subsection{Results: Some KST parameters which have transients which amplify by $10^{32}$}
\begin{figure}
\includegraphics{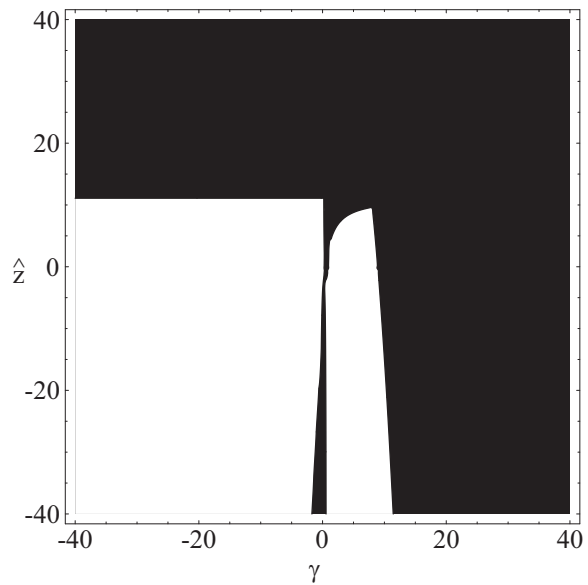}
\caption{\label{fig:pg:net} The shaded region indicates those KST parameters which produce
  some radially-propagating prototypical coherent wave packets which amplify their energy by
  greater than $10^{32}$ within the future domain of dependence of the slice $r\in[2,10]$. 
  Note a large proportion of parameter space has been excluded.
}
\end{figure}

Under the proper choice of KST parameters, shown shaded in Fig. \ref{fig:pg:net}, one of the three types of ray ($s=\pm 1,0$)  may
admit some prototypical coherent wave packet of 
polarization $\mu$ whose energy amplifies by $10^{32}$ [i.e. ${\cal A}_{s\mu} \ge 10^{32}$].
  The clear region  in Fig. \ref{fig:pg:net} indicates KST parameters for which we have not
 yet found a transient which amplifies by $10^{32}$.

\subsection{\label{sec:sub:pg:Generalize}Generalizations of our method which could generate stronger constraints on KST parameters}
With our  \emph{extremely} conservative approach -- eliminating those formulations with
wave-packet solutions which amplify by $10^{32}$ in the future domain of dependence -- we have
already eliminated a broad region of parameter space.
By relaxing some of our very
restrictive assumptions, we expect we could discard still more KST parameters:
\begin{enumerate}
\item \emph{Lower amplification cutoff}:
 Currently, we require an \emph{enormous}
amplification before we eliminate a formulation; relaxing the requirement on amplification
excludes more systems.  

\item \emph{Consider more transients}
Currently, we compute the amplification of only  a handful of transients; a consideration of other
transients (for example, in the neighborhood of circular photon orbits in PG) may allow us to
exclude additional parameters.

\item \emph{Consider a larger region}
Currently,  we limit attention only to
the future domain of dependence of the initial data slice.  Certain rays, however, remain
within the computational domain for far longer.  For example, in the PG case, rays near the
horizon remain in the domain for arbitrarily long;\footnote{One must take care to use the rays
  near the horizon in a sensible fashion.  While analytically the rays remain within the
  computational domain for arbitrarily long times, one cannot expect wave packet
  solutions to be resolved and present in a numerical solution for arbitrarily long:
  the code has a finite smallest resolved scale.  In practice, one must remember that whatever
  amplification one computes must be realistically attainable by some numerical simulation of
  fixed (though perhaps high) resolution in the coordinates of interest.} 
even the slowly-infalling rays last
substantially longer than the domain of dependence.  Therefore, by considering the
amplification of transients over a longer interval, we will discover significantly greater
amplification and thus exclude a significantly broader class of formulations of Einstein's
equations.

\item \emph{Combine with boundary conditions}  
Finally, if we determine how geometric-optics solutions interact with boundary conditions, we
can generalize our approach and address the \emph{late-time stability} properties of the
evolution equations -- or, in other words, address the stability properties of the full
initial-plus-boundary value problem. 
\end{enumerate}

\section{Conclusions}
In this paper, we have demonstrated that  certain transients (prototypical coherent
wave packets) can be used to veto a significant range of proposed formulations of Einstein's
equations.    We have described in considerable pedagogical detail precisely how to construct
expressions  for  (or estimates of)
the growth rate of prototypical  coherent wave packets
[i.e. Eq. (\ref{eq:rate:prototypical})], verify those estimates, and employ them to veto 
proposed formulations of Einstein's equations.    These expressions employ no free parameters
or knowledge of the solution, aside from a choice of plausible rays to examine.   Moreover,
despite the sometimes exhaustive details provided in Sections
\ref{sec:KSTtests:Rindler} and \ref{sec:KSTtests:PG}, the key tool -- the growth rate of
prototypical coherent wave packets [Eq. (\ref{eq:rate:prototypical})]
-- is easy to apply, with little conceptual, notational, or computational overhead (see, for
example, the brief Sec. \ref{sec:sub:pg:slow:quickway} and its application in
Sec. \ref{sec:KSTnumerics:pg}).
Whether they are used conservatively, as in this paper, or generalized along the lines
suggested in Sec. \ref{sec:sub:pg:Generalize} (i.e. using more rays and larger fragments of
spacetime), 
we believe these techniques will provide a
useful way to bound the number of proposed formulations before further tests are conducted
 (for example  by the more ambitious Lindblom-Scheel energy-norm method) to decide
whether a given formulation can produce effective simulations.  

While our the specific examples of  analyses in this paper have employed linearizations of the
field equations 
themselves, we could just as well linearize a FOSH system representing evolution equations for
the constraint fields [see, for example, KST Eqs. (2.40-2.43)].    The evolution equations for
the 
constraints have been emphasized by many other authors as an 
probe of unphysical behavior.   Since the general arguments of
Sections \ref{sec:rays}, \ref{sec:packets}, and \ref{sec:energy} do not depend on the precise
FOSHLS used, we can perform a calculation following the same patterns as (for example)
Sec. \ref{sec:KSTnumerics:pg} to discover ill-behaved formulations.\footnote{The author expects no new
information can be obtained by such an analysis.  Moreover, because the constraint equations,
when written in first-order form,
involve many more variables than the field equations themselves, such an analysis should prove substantially more
challenging. 
}

In this paper,  we have also discovered curious
properties of modes trapped on the horizon of a Schwarzchild hole in PG coordinates
(Sec. \ref{sec:KSTnumerics:pg}).  Analytically, we would expect that, if any growth rate for
modes trapped on the horizon were
positive, then these modes
should grow without bound and be present in the evolution at late times.  Numerically, however,
we know that no \emph{resolved} wave packets can appear at late times: such solutions would
have to initiate arbitrarily close to the horizon, inconsistent with  resolved,
finite-resolution initial data.  Still,
marginally-resolved solutions of similar character could potentially behave in an
implementation-dependent fashion, seeding outgoing modes which then propagate and amplify into
the domain for all time.  We shall explore this possibility in a future paper.

\appendix

\section{\label{ap:identities}Useful identities used in the text}

\subsection{An alternative approach to the group velocity}
The eigenequation which defines the natural polarization spaces $B_j$
and their associated eigenvalues $\omega_j$,
\[
A^a p_a v_{j,\alpha} = \omega_j v_{j,\alpha} 
\]
for each $v_{j,\alpha}\in B_j$ (cf. Sec. \ref{sec:sub:defs}),  may be alternatively expressed as 
\begin{equation}
A^a(x) p_a P_j(x,p) = \omega_j(x,p) P_j(x,p) \; .
\end{equation}
for $P_j$ the projection operator to the $j$th eigenspace of $A^a p_a$.
Differentiate this expression relative to $p_b$, then apply $P_j$ from the left, to find
\begin{equation}
\label{eq:tool:groupV}
P_j A^a P_j = \frac{\partial \omega_j}{\partial p^a} P_j  = V_j^a  P_j\; .
\end{equation}
Equivalently, if $v_{j,\alpha}$ are a collection of  basis
vectors for the $j$th eigenspace which are orthonormal relative to the inner product generated
by $S$, 
\begin{equation}
\label{eq:tool:groupV2}
(v_{j,\beta}, S A^a v_{j,\alpha}) = \delta_{\alpha \beta} V_j^a \; .
\end{equation}

\subsection{\label{ap:sub:norotate}The no-rotation condition}
In Section \ref{sec:sub:polarizeSimplify}, we claim we can always find a basis for $B_j$ at
each point in the neighborhood of a given ray which satisfies the \emph{no-rotation condition}
[Eq. (\ref{eq:trial:PolarizationBasisPropagate})]:
\begin{equation}
\label{eq:norot2}
   \left(v_{j,[\alpha}, S(\partial_t + A^a \partial_a) v_{j,\beta]}   \right) =0
\end{equation}
for all $\alpha$, $\beta$.   In this section, we demonstrate explicitly how to construct a
basis which satisfies both the no-rotation condition and which remains orthonormal.

  If the right-hand side term in  Eq. (\ref{eq:norot2}) is
not zero in the basis $v_{j,\alpha}$, we attempt to choose a new basis 
\[
v_{j,\bar{\alpha}} \equiv  R_{\bar{\alpha}\alpha} v_{j,\alpha}
\]
such that Eq. (\ref{eq:trial:PolarizationBasisPropagate}) is satisfied by the new basis and
moreover such that the new basis is orthonormal.  In particular, if the no-rotation condition
is  satisfied in the new basis, then (because $P_j A^a P_j = V_j^a$) we know
\begin{eqnarray}
0&=& \sum_{\alpha  \beta} R_{\bar \alpha \alpha} R_{\bar \beta \beta}  \left(v_{j,[\alpha}, S(\partial_t + A^a \partial_a) v_{j,\beta]}   \right) \\
 &+&R_{[\bar{\alpha}\alpha} \sum_\alpha (\partial_t + V_j^a \partial_a)  R_{\bar{\beta}]\alpha}
 \; . \nonumber
\end{eqnarray}
[In the second term above, the antisymmetrization is over only the barred indicies $\bar \alpha$ and $\bar
\beta$.]  On the other hand, if the basis is orthonormal, then  $\sum_\alpha R_{\bar{\alpha}\alpha}
R_{\bar{\beta}\alpha} = \delta_{\bar{\alpha}\bar{\beta}}$, implying 
\[
  \sum_\alpha R_{(\bar{\alpha}\alpha} (\partial_t + V_j^a \partial_a)  R_{\bar{\beta})\alpha}
  = 0 \; .
\] 
[In the above, the operator is symmetrized over the indicies $\bar \alpha$ and $\bar \beta$.]
Therefore, combining the two, we conclude that if the new basis is both orthonormal and
satisfies the no-rotation, the matrix 
$R$ must satisfy  the ordinary differential equation
\[
(\partial_t + V_j^a \partial_a) R_{\bar{\alpha} \alpha}
  = - 
   \left(v_{j,[\alpha}, S(\partial_t + A^a \partial_a) v_{j,\beta]}   \right)
    R_{\bar{\alpha}\beta}
\]
subject to initial data $R_{\bar{\alpha}\alpha}=\delta_{\bar{\alpha}\alpha}$.  Solutions for
$R$ and thus $v_{j,\bar{\alpha}}$ exist in the neighborhood of a ray.  

\subsection{\label{ap:sub:reorg}Reorganizing inner products for the polarization equation}
In this section, we describe how to rearrange matters so the last term in Eq. (\ref{eq:trial:PolarizationConfusing}) -- namely, 
\begin{eqnarray}
\label{eq:simp:starthere}
   \bigg( v_{j,\alpha},\; 
  S(\partial_t + A^a \partial_a )v_{j,\beta} \bigg) \; 
\end{eqnarray}
-- has a simpler form.
If we choose a basis that satisfies the no-rotation condition
[Eq. (\ref{eq:trial:PolarizationBasisPropagate})], the antisymmetric part of this matrix is
zero.  Further, we may express the symmetric part of this expression by using the relations
\begin{eqnarray}
\partial_t (v_{j,\alpha}, S v_{j,\beta}) &=& 
(v_{j,\alpha}, S {\partial}_t
v_{j,\beta})  \\&&
 +  (v_{j,\beta}, S {\partial}_t v_{j,\alpha}) \nonumber\\&&
 + \left(v_{j,\alpha}, (\partial_t S) v_{j,\beta}\right) \nonumber \\
\partial_a (v_{j,\alpha}, S A^a v_{j,\beta}) &=& 
(v_{j,\alpha}, S A^a {\partial}_a v_{j,\beta}) \nonumber \\&&
 +  (v_{j,\beta}, S A^a {\partial}_t v_{j,\alpha}) \\&&
 + \left(v_{j,\alpha}, \left(\partial_a SA^a\right) v_{j,\beta}\right) \nonumber
\end{eqnarray}
[where we have observed that $SA^a$ and $A^a$ are symmetric relative to the canonical inner
product] and the expressions
\begin{eqnarray}
 (v_{j,\alpha}, S v_{j,\beta}) &=& \delta_{\alpha \beta} \\
 (v_{j,\alpha}, S A^a v_{j,\beta}) &=& \delta_{\alpha \beta} V_j^a
\end{eqnarray}
[i.e. orthogonality and Eq. (\ref{eq:tool:groupV2})].   These relations tell us that, if the
no-rotation condition is satisfied, 
\begin{eqnarray}
&\left(v_{j,\alpha}\right. ,&  \left.S(\partial_t + A^a\partial_a) v_{j,\beta}\right)
= \\
 & &  \frac{1}{2}\delta_{\alpha,\beta} \partial_a V^a(\vec{x},\vec{k}(x))
 - \frac{1}{2}\left(v_{j,\alpha}, \left(\partial_t S+ {\partial}_a S A^a\right) v_{j,\beta}\right) \nonumber 
\end{eqnarray}
Our notation for the first term on the right side (i.e. the divergence of the group velocity)
is chosen to  emphasize that the derivative $\partial_a$ acts on \emph{all} the dependence on $\vec{x}$ -- in
particular, on any variation of $k_a = \partial_a \phi$ with $\vec{x}$.

\section{\label{ap:ValidityOfRayOptics}Demonstrating that the ray-optics approach provides
  high-quality approximate solutions to the FOSHLS}

For any fixed initial data, the ray-optics solution obtained in Sec. \ref{sec:rays} will break down at some
point along each  ray.  In this section, we estimate how long a solution obtained by solving
Eq. (\ref{eq:trial:Net}) can be trusted.

Specifically, in Sec. \ref{ap:est:rewrite} we express the FOSHLS [Eq. (\ref{eq:def:foshl})]
using alternative variables better-suited to describing the geometric-optics solution.  Next,
in Sec. \ref{ap:est:scales} we survey the various orders of magnitude that arise in the
problem.  Using those orders of magnitude, in Sec. \ref{ap:est:errors} we estimate the error in
Eq. (\ref{eq:def:foshl}) that occurs when a geometric optics solution is substituted for $u$
(e.g. we estimate how close the norm of the left side is to zero).  Finally, knowing how much
error we make when using a geometric-optics solution, in Sec. \ref{ap:est:trust} we estimate
how errors involved in a geometric-optics approximation grow; we therefore discover how long we
can trust a purely geometric-optics-based evolution.

\subsection{\label{ap:est:rewrite}Review: Writing equations in terms of $\phi$ and $d_{l,\alpha}$}
In Equations (\ref{eq:defImplicit:ub}) and (\ref{eq:defImplicit:ds}) we describe how to
parameterize the $N$-dimensional state vector $u$ by using $N$ functions $d_{l,\alpha}$ and one
additional 
function $\phi$.  If we insert this substitution into the original FOSHLS
[Eq. (\ref{eq:def:foshl})], then dot the results against each of the 
orthonormal basis vectors $v_{l,\alpha}$, we obtain the equations
\begin{eqnarray}
0&= &i (v_{l,\alpha},S \bar{u})\left[ \partial_t + V_{l}^a\partial_a\right] \phi \nonumber \\
  & & +\bigg( v_{l,\alpha},\; S (\partial_t + A^a \partial_a - F)\bar{u} \bigg) \nonumber  \\
 &=& i d_{l,\alpha}\left[ \partial_t + V_{l}^a\partial_a\right] \phi  \\
 & &+  \partial_t d_{l, \alpha} + \sum_m \sum_\beta(v_{l,\alpha},\; S A^a v_{m,\beta}) \partial_a d_{m,\beta} \nonumber\\
  & & + \sum_m \sum_\beta d_{m,\beta}\bigg( v_{l,\alpha},\; S(\partial_t + A^a \partial_a -
  F)v_{m,\beta} \bigg) \; . \nonumber  
\end{eqnarray} 
In the above, we have observed that $A^a$ is symmetric relative to the inner product generated
by $S$  and that
$v_{j,\alpha}$ is an 
eigenvector of $A^a \partial_a \phi$  with eigenvalue  $\omega_j = V^a_{j} \partial_a \phi$.  We
can further reorganize this equation by pulling out 
all terms that involve $d_l$ explicitly, and also  by using Eq. (\ref{eq:tool:groupV}) to
simplify $(v_{l,\alpha}, S A^a v_{l,\beta})=V_l^a \delta_{\alpha\beta}$:
\begin{eqnarray}
\label{eq:generalFOSHL}
0 &=& i d_{l,\alpha}\left[ \partial_t + V_{l}^a\partial_a\right] \phi +  \left[\partial_t
   +  V_l^a \partial_a \right] d_{l,\alpha}
   \\
  & &  + \sum_\beta d_{l,\beta}\bigg( v_{l,\alpha},\; S (\partial_t + A^a \partial_a - F)v_{l,\beta} \bigg) \nonumber 
   \\
 & &+ \sum_{m\ne l} \sum_\beta(v_{l,\alpha},\; S A^a v_{m,\beta}) \partial_a d_{m,\beta} \nonumber\\
  & & + \sum_{m\ne l} \sum_\beta d_{m,\beta}
  \bigg( v_{l,\alpha},\; S(\partial_t + A^a \partial_a
   - F)v_{m,\beta} \bigg) \; . \nonumber  
\end{eqnarray} 

\subsection{\label{ap:est:scales}Natural scales used in order-of-magnitude estimates}
To make order-of-magnitude arguments regarding the
solution, we need to understand how the natural length scales of the problem enter into it.

Rather than complicate the order-of-magnitude calculation unnecessarily, we shall for
simplicity proceed as if there existed only one characteristic speed. 
In other words, we shall freely convert between space and time units by using the norm of
$A^a$; for example, we can interpret $\tau_{F,n}|A|$ as a  natural length scale.  Finally, for
brevity, we shall assume space and time units are chosen so $|A|\sim 1$.

Even with the above simplification, many natural scales arise in the problem, including the magnitude of
$F$; the natural length and time scales on which $F$ and $A$ vary; and the length scale on
which the 
initial data varies.  Again, for simplicity we shall summarize all these scales by only two numbers:
\begin{itemize}
\item \emph{``Length'' scale} ($L$)  We define the natural ``length'' scale to be the natural
  time scale that enters on the right side of Eq. (\ref{eq:trial:PolarizationConfusing}).  To
  be explicit, $L$ is the smaller of $|d|/|A| |\partial_a d|$ and $1/|F|$.
\item \emph{``Variation'' scale} ($\tau_\text{vary}$)  The remaining scales do not arise
  directly in the equation.  They affect the propagation of the wave packet only because they
  determine the rate at which terms in the equation are modulated as the wave packet propagates
  in space and time.  We therefore call the smallest of the remaining scales the
  \emph{variation scale} ($\tau_\text{vary}$); its value is the smallest of the length and
  time scales on which $A$ and $F$ vary.
\end{itemize}

\subsection{\label{ap:est:errors}Degree to which ray-optics solution satisfies the FOSHLS}

Using the above rough estimates ($L$ and $\tau_\text{vary}$) to characterize the magnitude of terms both used and
neglected,  we find that geometric-optics solutions [Eq. (\ref{eq:trial:Net})] very nearly
satisfy the full  FOSHLS [Eq. (\ref{eq:def:foshl}), alternatively expressed as Eq. (\ref{eq:generalFOSHL})].
To be explicit, when we insert a geometric-optics solution which propagates in the $j$th polarization [i.e. a
solution to Eq. (\ref{eq:trial:Net})] into Eq. (\ref{eq:generalFOSHL}), we find the following:
\begin{eqnarray}
0 &=& i d_{l,\alpha}(\omega_l  -\omega_j) 
  +\left[\partial_t  +  V_l^a \partial_a \right] d_{l,\alpha}
   \\
  & &  + \sum_\beta d_{l,\beta}\bigg( v_{l,\alpha},\; S (\partial_t + A^a \partial_a - F)v_{l,\beta} \bigg) \nonumber 
   \\
 & &+ \sum_{m\ne l} \sum_\beta(v_{l,\alpha},\; S A^a v_{m,\beta}) \partial_a d_{m,\beta} \nonumber\\
  & & + \sum_{m\ne l} \sum_\beta d_{m,\beta}
  \bigg( v_{l,\alpha},\; S(\partial_t + A^a \partial_a
   - F)v_{m,\beta} \bigg) \nonumber  
\end{eqnarray} 
[Here, we have used Eq. (\ref{eq:trial:Phase}) and the definition of $\omega_j$ to simplify
the first term.] 

We have two circumstances:
\begin{itemize}
\item When $l=j$, the first three terms precisely cancel [see Eqs. (\ref{eq:trial:Phase}) and
(\ref{eq:trial:PolarizationConfusing})].  The only terms remaining are of order $d_{m,\beta}$ 
$m\ne j$.
\item On the other hand, when $l\ne j$, the first term does not cancel.  Rather, it is large,
  because $\partial_a \phi$ is large (i.e. we are in the short-wavelength limit), and the
  $\omega_l$ are proportional to $\partial_a \phi$.    

For brevity, assume the eigenvalues of
  $A^a$ are of comparable magnitude, so to order of magnitude 
  $\omega_j \sim \omega_l \sim\omega \sim \omega_j  - \omega_l$.  We may then express the
  equation when $l \ne j$ as
\[
0= {\cal O}(\omega d_{l\beta}) + {\cal O}(d_{j,\alpha}/L)
\]
The second terms will force the first terms, generally, to be nonzero. 
\end{itemize}
From the second case, we know that when  $l\ne j$, $|d_l|\sim {\cal O}(|d_j|/L\omega)$.
Combining this result with the first equation, we conclude that when we use our trial solution,
we are ignoring terms of order $|d_j|/L^2 \omega$ 
 when $l=j$ and terms of order and $|d_j|/L$ when $l\ne j$.

\subsection{\label{ap:est:trust}Length of time ray-optics solution can be trusted}
To estimate the integrated effects the neglected terms have on the diagonal and off-diagonal
polarization amplitudes $d_{j,\alpha}$ and $d_{l,\alpha}$, respectively, we integrate the
previous equations.

When $l\ne j$, we have a DE of form
\[
\frac{d}{ds} d_{l,\beta} + i d_{l,\beta} \Delta \omega  + {\cal O}(|d_j|/L) = 0
\]
where we neglect smaller terms, where $d/ds = \partial_t + V^a_j \partial_a$ represents the
derivative along a characteristic, 
and where $\Delta \omega = \omega_l - \omega_j\sim \omega$.
Since we limit attention to $\omega$ very large (i.e. $\omega \tau_\text{vary} \gg 1$), we may ignore the weak effects of any time
variation of $L$ and treat it as constant. 
Since
$|d_j|$ varies along the characteristic much
more slowly than does $\exp(-i\omega s)$, we find that after an affine length $T$, 
$|d_l|$ will be of order 
\begin{equation}
|d_l| \sim T |\partial_s d_j|/\omega L \sim |d_j| T /L^2 \omega \; .
\end{equation}
(Here, I assume $L^2\omega$ is suitably averaged, as
 $L$ will vary as the path evolves.)
Similarly, when $l=j$, we  ignore terms of order $|d_j|/L^2 \omega$.  We have a DE of form
\[
\frac{d}{ds} d_{j,\alpha} + (\text{known, real}) + {\cal O}(|d_j|/L^2 \omega)=0
\]
Therefore, integrating along an affine length $T$ of the  ray, we
expect errors in the $d_{j\alpha}$'s
of relative magnitude 
\begin{equation}
\label{eq:def:diffractionError}
 \epsilon_\text{amp}= T/L^2 \omega
\end{equation}
 when $\epsilon_\text{amp}$ is sall.  
In both cases, we see  the neglected terms will be smaller than $|d_j|$ by magnitude $\epsilon_\text{amp}$.

If we are simulating a \emph{given} system, with fixed initial data, we can only trust a
solution out to time 
$T\sim L^2 \omega$.  
However, for any compact region of any characteristic (i.e. for
any fixed $T$), we can always choose $\omega$ sufficiently large so the relative errors $\epsilon_\text{amp}$ is
arbitrarily small.

\section{\label{ap:prototypesAsGeneric}When do PCWPs exist?}
Rather than evolve general wave packets using the full geometric-optics equations, for simplicity in
this paper we often restrict attention to prototypcial coherent wave packets (PCWPs), which --
if they exist -- vastly simplify the problem of evolution
(cf. Sec. \ref{sec:sub:packets:prototypical}).    In this appendix, we try to clarify the
conditions under which prototypical coherent wave packet solutions exist as exact or approximate
solutions to the geometric-optics equations.

We can better understand under what conditions prototypical coherent wave packets exist if we
rewrite the  polarization  equation
[Eq. (\ref{eq:trial:PolarizationEasy})] using the basis $f_{j}^{(\mu)}$
[Eq. (\ref{eq:def:fj})].   When we do so, we find PCWP are exact solutions only for special circumstances.
However, when some eigenvalue of $O_j$ is large, we find that PCWPs arise naturally as limits
of \emph{arbitrary} coherent wave packets.

\subsection{Rewriting polarization equation in the basis of eigenvectors of $O_j$}
\emph{Basis vectors and their components}: The basis vectors $f_j^{(\mu)}$ are defined above.
Since we express the polarization equation in component form relative to some no-rotation
basis, we also need notation for the components $f_{j\alpha}^{(\mu)}$ of these basis vectors relative to the
no-rotation basis:
\[
f_j^{(\mu)} = \sum_\alpha f_{j\alpha}^{(\mu)} v_{j,\alpha} \; .
\]

\emph{Dual vectors and their components}: The basis vectors $f_j^{(\mu)}$ are not necessarily
orthogonal.  To facilitate computations, we define a dual basis $\tilde{f}_j^{(\mu)}$ such that 
\[
\delta^{\mu \nu} 
 =\left( \tilde{f}_j^{(\nu)}, \; S f_j^{(\mu)}\right) \; .
\]
The dual basis vectors  can be expressed in terms of components, denoted
$\tilde{f}_{j,\alpha}^{(\mu)}$, relative to the 
no-rotation basis.

\emph{Explicitly rewriting polarization equation}: Substituting in the expansion 
\[
\bar{u}=\sum_\mu D_{j\mu} f_j^{(\mu)} \quad \leftrightarrow \quad
d_{j\alpha} = \sum_\mu D_{j\mu} f_{j\alpha}^{(\mu)}
\]
into the polarization equation [Eq. (\ref{eq:trial:PolarizationEasy})], then using the dual
vector basis to select specific components, we find
\begin{eqnarray}
\label{eq:tool:polarizationPrototypicalEquation}
0
 &=&\left(\partial_t + V_j^a\partial_a 
   + \frac{1}{2} \partial_a V_j^a - o_{j\mu}\right) D_{j\mu} \\
 & & - \sum_\nu D_{j\nu} \left[\sum_\alpha  
  \bigg(\tilde{f}_{j\alpha}^{(\mu)}, \; S
   \left(  
     \partial_t + V_j^a \partial_a 
   \right) f_{j\alpha}^{(\nu)} \bigg ) \right]
  \nonumber
\end{eqnarray}

\subsection{Sufficient conditions for PCWP to be exact solution}
By definition, a prototypical coherent wave packet solution associated with the polarization
direction  $f_{j}^{(\mu)}$ exists only if there is a solution -- exact or approximate -- to Eq. 
(\ref{eq:tool:polarizationPrototypicalEquation})
with all $D_{j\nu}=0$ except for $\nu=\mu$ [i.e. $D_{j,\mu}\ne 0$].  In other words, for a
complete collection of PCWP solutions to exist, one for each $\mu$, the third
term in Eq. 
(\ref{eq:tool:polarizationPrototypicalEquation}) must be diagonal, or zero.

[In fact, for the examples addressed in this paper (i.e. in Sections
 \ref{sec:KSTtests:Rindler} and \ref{sec:KSTtests:PG}, for propagation on the light cone), the
 third term is in fact exactly zero.]

\subsection{PCWP as limit of arbitrary rapidly-growing coherent wave packet}
If the largest eigenvalue $o_{j\nu}$ of $O_j$ is particularly large compared to the third
term, then generic solutions to the polarization equation 
 [Eq. (\ref{eq:tool:polarizationPrototypicalEquation})] 
will converge to a state with $D_{j\nu}\gg D_{j\mu}$ for $\mu \ne \nu$ (i.e. $w=f_j^{(\nu)}$).
In other words, if the largest eigenvalue of $O_j$ is large, then generic wave packets will
converge to the PCWP with $w=f_j^{(\nu)}$.

\section{\label{ap:energybound}Bounding the energy growth rate}
In Sections \ref{sec:rays} and \ref{sec:packets} we introduced wave-packet solutions as
solutions which are localized in the neighborhood of a given ray.  However, while we obtaind
expressions for the growth rate of certain specialized wave packet solutions
[Sec. \ref{sec:energy}], in the main text of this paper we never provided a strict bound on the
growth rate of a wave packet.

In fact,  we can bound the instantaneous growth rate of a coherent wave packet
[Eq. (\ref{eq:rate:coherent})]
by a quantity independent of the precise polarization state $w$ of that packet:
\begin{equation}
\frac{1}{E} \frac{dE}{dt} \le \text{max}_{w \in B_j} \frac{(w ,S Q w)}{(w, S w)}
\end{equation}
As $Q$ is symmetric relative to $S$, it has a spectrum of real eigenvalues, each associated
with  eigenspaces that are orthogonal relative to $S$.  It follows that
if $\kappa_s$ is the largest eigenvalue of $Q$,
\begin{equation}
\label{eq:result:weakbound}
\frac{1}{E} \frac{dE}{dt} \le  \kappa \; .
\end{equation}
This procedure follows precisely the same outline as the energy-norm upper bound discussed in
LS Eq. (2.17) and (2.18).

This upper bound on the growth rate for all polarizations propagating along a given ray can be
used as a line-by-line replacement for the maximum growth rate of PCWPs
[Eq. (\ref{eq:rate:prototypical})] in practical calculations to determine the largest
amplification possible by a wave packet propagating in the future domain of dependence
[e.g. Sections  \ref{sec:KSTnumerics:rindler} and \ref{sec:KSTnumerics:pg}]

\section{\label{ap:KST}Rays optics and KST 2-parameter formulation}
While KST introduce a very large family of symmetric hyperbolic systems, they emphasize
(and limit their calculations to) a simple 2-parameter subset.  This two parameter system has
both  physical characteristic speeds and a simple principal part (i.e. simple form for $A^a$).
As a result, the algebra required for its ray-optics limit (i.e. computations of $\omega_j$,
etc) proves particularly simple.

\subsection{Generally}
The KST system has as variables the tensors $g_{ab}$, $P_{ab}$, $M_{kab}$ defined over 3-space,
for a total of 6+6+18=30 fields.

\subsubsection{Principal part and symmetrizer}
The principal part has the form [KST Eq. (2.59), along with the definition of
$\hat{\partial}_o$ in KST  Eq. (2.10)]:
\begin{subequations}
\label{eq:pp}
\begin{eqnarray}
\label{eq:pp:1}
(\partial_t - \beta^a \partial_a) g_{ij}  &\simeq& 0 \\
\label{eq:pp:2}
(\partial_t - \beta^a \partial_a) P_{ij} + N g^{ab} \partial_a M_{bij} &\simeq& 0 \\
\label{eq:pp:3}
(\partial_t - \beta^a \partial_a) M_{kij} + N \partial_k P_{ij} &\simeq& 0 
\end{eqnarray}
\end{subequations}
After linearizing about a background solution, this principal part and a choice of
representation for the fields (i.e. $u$) gives us the explicit form for $A^a$.  We may
represent the result as
\begin{equation}
\label{eq:def:AexpandKST}
A^a = -\beta^a \textbf{1} + N A^a_o
\end{equation}
for \textbf{1} the identity operator and $A_o^a$ a construction which depends only on the
background metric $g$ and the choice of field ordering used in going to a matrix representation
(i.e. the order of the fields in $u$).

This principal part is symmetric hyperbolic, using as symmetrizer (for example)
LS Eq. (3.67):
\begin{eqnarray}
\label{eq:def:symmetrizerKST}
(u,Su) = g^{a\bar{a}} g^{b\bar{b}} dg_{ab} dg_{\bar{a}\bar{b}} 
  +  g^{a\bar{a}} g^{b\bar{b}} dP_{ab} dP_{\bar{a}\bar{b}} \nonumber\\
  +
   g^{a\bar{a}} g^{b\bar{b}} g^{k\bar{k}} dM_{kab} dM_{\bar{k}\bar{a}\bar{b}} \; .
\end{eqnarray}
This symmetrizer (represented here in LS notation) amounts to nothing more than the
naturally-constructed sum of squares of components of $g$, $P$, and $M$.

\subsubsection{Eigenvalues and group velocity}
From the principal part, we can deduce the three possible eigenvalues:
\begin{eqnarray}
\label{eq:result:KSTgeneralOmegas}
\omega_s(x,p) &=& -\beta^a p_a + s N \sqrt{g^{ab}p_a p_b}
\end{eqnarray}
where $s =0,\pm 1$.  From this expression we obtain the group velocities
\begin{eqnarray}
\label{eq:result:KSTgeneralVgroup}
v_s^a(x,p) &=& -\beta^a  + s N \hat{p}^a
\end{eqnarray}
where
$
\hat{p}^a \equiv g^{ab}p_b / \sqrt{g^{rs}p_r p_s}
$.

\subsubsection{\label{sec:sub:basis}Eigenfields and projection operators}
KST tabulate the eigenfields of the principal-part operator
[Eq. (\ref{eq:pp})] in KST Eq. (2.61) and the surrounding text.  These
expressions yield the following orthonormal basis vectors for the three eigenspaces of $A^a
\hat{p}_a$:
\begin{subequations}
\label{eq:basis}
\begin{eqnarray}
\label{eq:basis:1}
v_{o,g,ab} &=& g_{ab} \\
\label{eq:basis:2}
v_{o,x,ab} &=& \frac{
  \left[M_{qab}\hat{x}^q -  \hat{p}_u \hat{x}^u \hat{p}^q M_{qab}\right]
  }{\sqrt{1-(p^a \hat{x}_a)^2}}
    \\
\label{eq:basis:3}
v_{o,y,ab} &=& \frac{
  \left[M_{qab}\hat{y}^q -  \hat{p}_u \hat{y}^u \hat{p}^q M_{qab} \right]
  }{\sqrt{1-(p^a \hat{y}_a)^2}}
   \\
\label{eq:basis:4}
v_{\pm,ab} &=& \frac{1}{\sqrt{2}}\left[P_{ab} \pm \hat{p}^q M_{qab}\right]
\end{eqnarray}
\end{subequations}
where we treat symbols for the
fields as  basis vectors in the space of fields, and 
where  $\hat{x}$ and $\hat{y}$ are two 3-vector fields not parallel to $\hat{p}$ and which
are  orthonormal relative to the metric $g_{ab}$ at each point.

\subsection{Special case: Flat spatial metric}
When the KST system is applied to a time-independent solution with a flat spatial metric, the
algebra simplifies  substantially.  For example, the symmetrizer $S$
[Eq. (\ref{eq:def:symmetrizerKST})] is  the 
identity operator on the set of fields.  The inner product
generated by $S$ is therefore constant in space and time.

\subsubsection{Simplifying general polarization equation}
Since we fix the basis vector convention by Eq. (\ref{eq:basis}), we must
use the polarization equation in the form of  Eq. (\ref{eq:trial:PolarizationConfusing})
(i.e. we do not generically expect the no-rotation condition to hold).   We therefore must
evaluate the term
\begin{eqnarray}
\left(v_{j,\beta}, S(\partial_t + A^a \partial_a ) v_{j,\beta} \right)  
 &=& \left(v_{j,\beta}, S(\partial_t - \beta^a \partial_a ) v_{j,\beta} \right) \nonumber \\
 & & + N \left(v_{j,\beta}, S A^a_o \partial_a  v_{j,\beta} \right) \nonumber
\end{eqnarray}
Since we know how the basis vectors change as a function of the congruence direction
$\hat{k}=\hat{p}$ [Eqs. (\ref{eq:basis:1}--\ref{eq:basis:4})], we can rewrite this expression
in terms of our knowledge of the congruence and $\beta^a$.

For example, for the fields propagating forward along the congruence at unit speed
($j=s=\pm 1$), we have
\begin{eqnarray}
\label{eq:tool:generalPolarize1}
\left(v_{s,ab}, S A^a_o \partial_a  v_{s,cd} \right)  =  \left(\frac{1}{2} s \partial_q \hat{p}^q  \right)
  \delta_{ac} \delta_{bd}\\
\label{eq:tool:generalPolarize2}
\left(v_{s,ab}, S  (\partial_t - \beta^a \partial_a ) v_{s,cd} \right)  = 0
\end{eqnarray}

\subsubsection{Simplifying the  energy equations}
The only new quantity needed to evaluate the  energy equation is  $\partial_a S A^a$
To evaluate $\partial_a S A^a$, we note $S$ is the identity, so we just differentiate the
result we obtain from Eq. (\ref{eq:def:AexpandKST}):
\[
\partial_a S A^a = -\partial_a \beta^a \textbf{1} + (\partial_a N) A^a_o
\]
Now, if we take the inner product over the fields, we arrive at the expression
\begin{equation}
\label{eq:tool:dSA}
(v_{j,\alpha},(\partial_a S A^a) v_{j,\beta}) = 
  \delta_{\alpha\beta}
  \left[
   -\partial_a \beta^a  + (\partial_a N) s_j  \frac{p^a}{|p|}
  \right]
\end{equation}

\begin{acknowledgements}
I thank Mark Scheel and Lee Lindblom for  valuable discussions, suggestions, and assistance
during the prolonged development of this paper.  This research has been supported in part by
NSF Grant PHY-0099568. 
\end{acknowledgements}

\end{document}